\documentclass[fleqn,usenatbib]{mnras}

\usepackage{newtxtext,newtxmath}
\usepackage[T1]{fontenc}

\DeclareRobustCommand{\VAN}[3]{#2}
\let\VANthebibliography\thebibliography
\def\thebibliography{\DeclareRobustCommand{\VAN}[3]{##3}\VANthebibliography}

\defcitealias{mun2024}{Paper I}

\usepackage{graphicx}	% Including figure files
\usepackage{amsmath}	% Advanced maths commands
\usepackage{orcidlink}  % Placing ORCID links for each author
\usepackage{hyperref}   % For hyperlinks to websites
\usepackage{caption}    % For placing subfigures
\usepackage{subcaption} % For placing subfigures
\usepackage{upgreek}    % Upright Greek symbols
\usepackage{xspace}

\newcommand{\Mstar}{$M_{\rm \star}$\xspace}
\newcommand{\Msun}{M\textsubscript{\sun}\xspace}
\newcommand{\Mhalo}{$M_{\rm halo}$\xspace}
\newcommand{\Mcrit}{$M_{\rm crit, 200}$\xspace}
\newcommand{\re}{$R_{\rm e}$\xspace}
\newcommand{\dsfr}{$\Delta$SFR\xspace}
\newcommand{\dssfr}{$\mathrm{\Delta \Sigma_{SFR}}$\xspace}
\newcommand{\sigstar}{$\mathrm{\Sigma_{\star}}$\xspace}
\newcommand{\sigsfr}{$\mathrm{\Sigma_{SFR}}$\xspace}
\newcommand{\sigstarunits}{\Msun~kpc\textsuperscript{-2}\xspace}
\newcommand{\sigsfrunits}{\Msun~yr\textsuperscript{-1}~kpc\textsuperscript{-2}\xspace}
\newcommand{\Halpha}{H$\upalpha$\xspace}
\newcommand{\eagle}{\textsc{EAGLE}\xspace}
\newcommand{\magc}{\textsc{Magneticum}\xspace}
\newcommand{\tng}{\textsc{IllustrisTNG}\xspace}
\newcommand{\simspin}{\textsc{SimSpin}\xspace}
\newcommand{\slopeunits}{dex $R_{\rm e}$\textsuperscript{-1}\xspace}
\newcommand{\dsfrsb}{\dsfr $>$ 0.5\xspace}
\newcommand{\dsfrms}{-0.5 $<$ \dsfr $<$ 0.5\xspace}
\newcommand{\dsfrgv}{-1.1 $<$ \dsfr $<$ -0.5\xspace}
\newcommand{\dsfrq}{\dsfr $<$ -1.1\xspace}

\title[Radial trends in MAGPI vs. simulations]{The MAGPI Survey: radial trends in star formation across different cosmological simulations in comparison with observations at $\boldsymbol{z \sim}$ 0.3}

\author[M. Mun et al.]{
Marcie Mun,$^{\orcidlink{0000-0002-3706-9955} 1,2}$\thanks{E-mail: jaeyeon.mun@anu.edu.au}
Emily Wisnioski,$^{\orcidlink{0000-0003-1657-7878} 1,2}$
Katherine E. Harborne,$^{\orcidlink{0000-0002-2043-7985} 3,2}$
Claudia D. P. Lagos,$^{\orcidlink{0000-0003-3021-8564} 3,2}$
\newauthor Lucas M. Valenzuela,$^{\orcidlink{0000-0002-7972-9675} 4}$
Rhea-Silvia Remus,$^{\orcidlink{0009-0008-9260-7278} 4}$
J. Trevor Mendel,$^{\orcidlink{0000-0002-6327-9147} 1,2}$
Andrew J. Battisti,$^{\orcidlink{0000-0003-4569-2285} 1,2,3}$
\newauthor Sara L. Ellison,$^{\orcidlink{0000-0002-1768-1899} 5}$
Caroline Foster,$^{\orcidlink{0000-0003-0247-1204} 6,2}$
Matias Bravo,$^{\orcidlink{0000-0001-5742-7927} 7}$
Sarah Brough,$^{\orcidlink{0000-0002-9796-1363} 6,2}$
Scott M. Croom,$^{\orcidlink{0000-0003-2880-9197} 8,2}$
\newauthor Tianmu Gao,$^{\orcidlink{0000-0002-1158-6372} 1,2}$
Kathryn Grasha,$^{\orcidlink{0000-0002-3247-5321} 1,2}$
Anshu Gupta,$^{\orcidlink{0000-0002-8984-3666} 9,2}$
Yifan Mai,$^{\orcidlink{0000-0003-3514-6280} 8,2}$
Anilkumar Mailvaganam,$^{\orcidlink{0009-0003-1221-1630} 10,11,2}$
\newauthor Eric G. M. Muller,$^{\orcidlink{0000-0001-5621-1577} 1, 2}$
Gauri Sharma,$^{\orcidlink{0000-0002-6070-2851} 12}$
Sarah M. Sweet,$^{\orcidlink{0000-0002-1576-2505} 13,2}$
Edward N. Taylor,$^{14}$
Tayyaba Zafar$^{\orcidlink{0000-0003-3935-7018} 10,2}$
\\
$^{1}$Research School of Astronomy and Astrophysics, Australian National University, Weston Creek, ACT 2611, Australia\\
$^{2}$ARC Centre of Excellence for All Sky Astrophysics in 3 Dimensions (ASTRO 3D), Canberra, ACT 2611, Australia\\
$^{3}$International Centre for Radio Astronomy Research, The University of Western Australia, 35 Stirling Highway, Crawley, WA 6009, Australia\\
$^{4}$Universitäts-Sternwarte, Fakultät für Physik,  Ludwig-Maximilians-Universität München, Scheinerstr. 1, 81679 München,  Germany\\
$^{5}$Department of Physics and Astronomy, University of Victoria, Finnerty Road, Victoria, British Columbia, V8P 1A1, Canada\\
$^{6}$School of Physics, University of New South Wales, Sydney, NSW 2052, Australia\\
$^{7}$Department of Physics \& Astronomy, McMaster University, 1280 Main Street W, Hamilton, ON L8S 4M1, Canada\\
$^{8}$Sydney Institute for Astronomy, School of Physics, University of Sydney, NSW 2006, Australia\\
$^{9}$International Centre for Radio Astronomy Research (ICRAR), Curtin University, Bentley WA, Australia\\
$^{10}$School of Mathematical and Physical Sciences, Macquarie University, NSW 2109, Australia\\
$^{11}$Macquarie University Astrophysics and Space Technologies Research Centre, Sydney, NSW 2109, Australia\\
$^{12}$Observatoire Astronomique de Strasbourg, Université de Strasbourg, CNRS UMR 7550, F-67000 Strasbourg, France\\
$^{13}$School of Mathematics and Physics, University of Queensland, Brisbane, QLD 4072, Australia\\
$^{14}$Centre for Astrophysics and Supercomputing, Swinburne University of Technology, John Street, Hawthorn, 3122, Australia\\
}

\date{Accepted XXX. Received YYY; in original form ZZZ}

\pubyear{2024}

\begin{document}
\label{firstpage}
\pagerange{\pageref{firstpage}--\pageref{lastpage}}
\maketitle

% Word limit = 250 words
\begin{abstract}
We investigate the internal and external mechanisms that regulate and quench star formation (SF) in galaxies at $z \sim 0.3$ using MAGPI observations and the \eagle, \magc, and \tng cosmological simulations. Using \simspin to generate mock observations of simulated galaxies, we match detection/resolution limits in star formation rates and stellar mass, along with MAGPI observational details including the average point spread function and pixel scale. While we find a good agreement in the slope of the global star-forming main sequence (SFMS) between MAGPI observations and all three simulations, the slope of the resolved SFMS does not agree within 1 -- 2$\sigma$. Furthermore, in radial SF trends, good agreement between observations and simulations exists only for galaxies far below the SFMS, where we capture evidence for inside-out quenching. The simulations overall agree with each other between $\sim1.5-4$ \re but show varying central suppression within $R \sim 1.5$ \re for galaxies on and below the SFMS, attributable to different AGN feedback prescriptions. All three simulations show similar dependencies of SF radial trends with environment. Central galaxies are subject to both internal and external mechanisms, showing increased SF suppression in the centre with increasing halo mass, indicating AGN feedback. Satellite galaxies display increasing suppression in the outskirts as halo mass increases, indicative of environmental processes. These results demonstrate the power of spatially resolved studies of galaxies; while global properties align, radial profiles reveal discrepancies between observations and simulations and their underlying physics. 
\end{abstract}


%%%%%%%%%%%%%%%%%%%%%%%%%%%%%%%%%%%%%%%%%%%%%%%%%%

%%%%%%%%%%%%%%%%% BODY OF PAPER %%%%%%%%%%%%%%%%%%

\section{Introduction}
\label{sec:intro}
In modern astronomy observations, theory, and simulations are commonly brought together to provide insight to the formation and evolution of galaxies. Observations provide snapshots of the real Universe at different times and wavelengths. These snapshots, and how they are processed, come with limitations complicating and sometimes biasing interpretations. Simulated universes provide a basis for testing theoretical models of galaxy formation and evolution for comparison to observational results. The combination of ingredients that shape a simulated universe is multi-faceted, where the choice of sub-grid physics, cosmological models, hydrodynamic techniques, etc., all contribute to the creation of a unique universe (e.g., \citealt{sd2015}, and references therein). Given these complexities, it’s unsurprising that different simulations can yield varying predictions for galaxy evolution. However, these discrepancies are beneficial, serving to refine and improve existing models and drive new observations.

Cosmological simulations are typically fine-tuned to reproduce the global properties of observed galaxies in the local Universe (i.e., $z \sim 0$; \citealt{genel2014, schaye2015, dubois2016, genel2018}). These include observables such as the stellar-to-halo mass, stellar-to-black hole (BH) mass, and mass-size relations, along with the stellar mass function and the cosmic star formation rate density. In contrast, resolved properties, e.g. star-formation gradients, age gradients, metallicity gradients, are rarely tuned serving as a good test of the multi-scale physics implemented in simulations. 

For example, observations in the local Universe indicate that galaxies show increasingly centrally suppressed SF with increasing stellar mass \citep{belfiore2018, ellison2018, bluck2020}, a strong indicator of inside-out quenching. On the other hand, the degree to which these trends are reproduced in simulations vary, some predicting either outside-in quenching \citep{starkenburg2019}, a mixture of both trends \citep{appleby2020}, or inside-out quenching \citep{nelson2021, mcdonough2023}.

These results highlight the necessity of pushing beyond investigating global galaxy properties to spatially resolving the effects of the different factors (e.g., various prescriptions of subgrid physics, numerics) on simulated galaxies. For example, the exact prescription of feedback from active galactic nuclei (AGN) has been shown to be crucial in reproducing the observed degree of central suppression of SF in passive galaxies, where X-ray and kinetic wind feedback have proven successful \citep{appleby2020, nelson2021}. Additionally, recent results also point towards the importance of robust measurements from observations, such as accounting for the sensitivity of the SFR and \Mstar measurements to the stellar population parameters used \citep{nelson2021}.

Given that the discrepancies seen in observations and simulations are not trivial to resolve, it is imperative to perform an `apples-to-apples' comparison to better constrain the key physics at play in observed galaxies. Previous studies have relied on raw output data from simulations, where they have attempted to reduce discrepancies by emulating the sample selection criteria from observations and `projecting' simulated galaxies by using face-on galaxies only. With the advent of integral field spectroscopy (IFS), an increasing number of studies have looked into building synthetic data cubes resembling observations of simulations \citep{vandesande2019, harborne2020, harborne2023, bh2022, nanni2022}. This trend is primarily motivated by the need to accurately mimic observing conditions (e.g., blurring due to atmospheric seeing) and account for the specific technicalities of the instruments used. 
 
One of the science goals of the Middle Ages Galaxy Properties with Integral Field Spectroscopy survey \citep[MAGPI\footnote{Based on observations obtained at the VLT of the European Southern Observatory (ESO), Paranal, Chile (ESO program ID 1104.B-0536)};][]{foster2021} is to build a set of mock data cubes from a suite of cosmological simulations for direct comparison to observations. The motivation behind this theoretical dataset lies in understanding which combination of different models for physical processes best reproduces observations, where the implementation of the effects due to observing conditions reduces the uncertainties in the comparison. 

In \citet{mun2024}, hereafter Paper I, we analysed how MAGPI galaxies regulate and quench their SF activity within 3 \re at $z \sim 0.3$. We found that star-forming galaxies above the star-forming main sequence (SFMS) have star formation enhanced throughout their discs, as manifested by a flat and elevated profile in SF. This suggests a mixture of physical processes at play (i.e., galaxy-galaxy interactions and efficient gas mixing of large gas reservoirs; \citealt{heckman1990, patton2013, davies2015, moreno2021}), whereas negative gradients are indicative of central starbursts likely caused by mergers \citep[e.g.,][]{thorp2019, ellison2020}. Galaxies on the SFMS also showed flat radial trends in SF, suggestive that these galaxies are subject to both short-term self-regulating processes \citep[e.g., outflows, cooling;][]{tacchella2016} and long-term variations \citep[i.e., dependent on halo mass assembly histories;][]{mattheeschaye2019}, the latter for which may not be reflected in use of SFR indicators sensitive to shorter timescales, such as \Halpha ($\sim$10 Myr, vs. $\sim$10 Gyr). On the other hand, galaxies below the SFMS in MAGPI showed positive gradients in SF radial trends, pointing towards mechanisms that preferentially quench galaxies from the inside-out. Given that MAGPI galaxies generally live in group environments, these results point towards a mixture of internal and external mechanisms, where the former dominates (e.g., AGN feedback, lower SF efficiencies in the galaxy centre; \citealt{wang2019, bluck2020, ellison2021, nelson2021, pan2024}) over the latter (e.g., ram pressure stripping, pre-processing; \citealt{schaefer2017, wang2022, brown2023}). 

In this study, we perform a follow-up investigation of the results presented in \citetalias{mun2024}, where we aim to disentangle the potential mechanisms that contribute to the overall shape of SF radial profiles. To expand upon \citetalias{mun2024}, we extend our analysis of MAGPI galaxies to mock data cubes from three simulations in the MAGPI theoretical library - \eagle \citep{schaye2015, crain2015}, \magc \citep{teklu2015, schulze2018}, and \tng \citep{naiman2018, springel2018, marinacci2018, nelson2018, pillepich2018b} \ -- to address the following questions: 1) do we reproduce the observed radial trends of MAGPI galaxies with mock data cubes from these simulations?; and 2) are environmental or internal processes dominant in shaping the radial trends in SF at this epoch?

This paper is structured as follows: we briefly introduce the MAGPI survey in Section \ref{sec:magpi} and the cosmological simulations used in Section \ref{sec:sims}. We then describe the methods used in Section \ref{sec:methods}, detailing the code used to generate mock data cubes, the various metrics used to quantify the degree of SF, and measurement of radial profiles. We present and compare radial profiles from MAGPI and all three simulations in Section \ref{sec:res}, then discuss the scientific implications in further detail in Section \ref{sec:disc}. We finally summarize our findings in Section \ref{sec:summ}. 

For the MAGPI sample, we assume a \citet{chabrier2003} initial mass function (IMF) and adopt a flat $\Lambda$CDM cosmology with $H_{0}$ = 70 km s\textsuperscript{-1} Mpc\textsuperscript{-1}, $\mathrm{\Omega_{m}}$ = 0.3, and $\mathrm{\Omega_{\Lambda}}$ = 0.7, as done in \citetalias{mun2024}. For the simulations, we adopt the cosmological parameters corresponding to each simulation.\footnote{There are slight differences in the cosmological parameters adopted in different simulations (see Section \ref{sec:sims}) which may affect the model-dependent measurements here, such as SFR and \Mstar. However, the differences are expected to be negligible, and as such we do not make any corrections based on the different cosmologies.} 

\begin{table*}
    \caption{Key information for each simulation used, where from left to right, we list the box run, hydrodynamics, simulation box size in cMpc, redshift of the snapshot used, average stellar particle mass in \Msun, initial gas particle/cell mass in \Msun, the final sample size of galaxies after SFR and \Mstar cuts on a spaxel-by-spaxel basis (as explained in Section \ref{subsec:sample}), the total number of central galaxies, and the total number of satellite galaxies.}
    \centering
    \begin{tabular}{cccccccccc}
    \hline
    Simulation & Box run & Hydrodynamics & Box size & $z$ & $\langle m_{\rm \star} \rangle$ & $m_{\rm gas}$ & N\textsubscript{gal} & N\textsubscript{cen} & N\textsubscript{sat} \\
     & & & (cMpc\textsuperscript{3}) &  & (\Msun) & (\Msun) & & &\\
    \hline
    \eagle & \textsc{Ref-L100N1504} & SPH (\textsc{gadget3}+\textsc{anarchy}) & 100\textsuperscript{3} & 0.271 & 1.8 $\times$ 10\textsuperscript{6} & 1.8 $\times$ 10\textsuperscript{6} & 8546 & 7550 (88\%) & 996 (12\%) \\
    \magc & \textsc{box4-uhr} & SPH (\textsc{gadget3}) & 68.2\textsuperscript{3} & 0.293 & 1.8 $\times$ 10\textsuperscript{6} & 1.0 $\times$ 10\textsuperscript{7} & 1382 & 984 (71\%) & 398 (29\%) \\
    \tng & TNG100-1 & MVM (\textsc{arepo}) & 110.7\textsuperscript{3} & 0.298 & 1.4 $\times$ 10\textsuperscript{6} & 1.4 $\times$ 10\textsuperscript{6} & 17721 & 11480 (65\%) & 6241 (35\%) \\
    \hline
    \end{tabular}
    \label{tab:sims_info}
\end{table*}

\section{The MAGPI Survey}
\label{sec:magpi}
Using the Multi-Unit Spectroscopic Explorer (MUSE) on the Very Large Telescope (VLT), MAGPI \citep{foster2021} is designed to study the spatially resolved stellar and ionised gas properties of galaxies at lookback times of 3 -- 4 Gyr ($0.25 < z < 0.35$). Combining the power of Ground Layer Adaptive Optics (GLAO) with MUSE has resulted in well-resolved maps of stellar and gas kinematics for hundreds of galaxies beyond the local Universe, which are directly comparable to those measured from IFS surveys such as the Mapping Nearby Galaxies at Apache Point Observatory \citep[MaNGA;][]{bundy2015} and the Sydney-AAO Multi-object Integral field spectrograph \citep[SAMI;][]{croom2012} surveys based at $z = 0$, in terms of relative spatial resolution. The sample consists of 60 primary/central galaxies with \Mstar $\mathrm{> 7 \times 10^{10}}$ \Msun and their neighbouring galaxies (`secondary' galaxies) residing in a broad range of environments (i.e., 12 $\mathrm{\lesssim \log}$ [\Mhalo/\Msun] $\lesssim$ 15) that were selected from the Galaxy and Mass Assembly Survey (GAMA; \citealt{driver2011, driver2022}). 

This study makes use of the star formation radial profile results from \citetalias{mun2024}, which were derived based on a total of 302 galaxies at $0.25 \leq z \leq 0.424$ with available data products, including emission line and stellar continuum fits, and stellar mass maps. Full details on the data reduction and emission/continuum fits are provided in Mendel et al. (in preparation) and Battisti et al. (in preparation), respectively. We refer the reader to Sections 2, 3.1, and 3.2 in \citetalias{mun2024} for descriptions of the above. 

Currently, the MAGPI theoretical library consists of \eagle, \magc, and \tng. Mock data cubes from each simulation are created using \simspin\footnote{\url{https://kateharborne.github.io/SimSpin/}} \citep{harborne2020, harborne2023}, a code that builds data cubes of individual simulated galaxies akin to those measured via IFS. \simspin is designed not only to be easily applicable to simulations built on varying hydrodynamic techniques, but also to match any observing instrument of choice. More details on the use of \simspin on the simulations in the MAGPI theoretical library will be provided in Harborne et al. (in preparation). We provide a brief description of the outputs used from \simspin in Section \ref{subsec:simspin}. 

\section{Simulations}
\label{sec:sims}
In this section, we briefly introduce each of the cosmological simulations used in the study. For comparison with the MAGPI data we use snapshots close in time to $z\sim0.3$ as indicated in Table \ref{tab:sims_info}. We list key technical details of the simulations and associated snapshots in Table \ref{tab:sims_info}.

\subsection{\eagle}
\label{subsec:eagle}
\eagle (Evolution and Assembly of GaLaxies and their Environments; \citealt{schaye2015, crain2015}) is a suite of cosmological hydrodynamical simulations, run on \textsc{anarchy} \citep{schaller2015}, a modified version of the $N$-body Tree-Particle-Mesh (Tree-PM) smoothed particle hydrodynamics (SPH) code \textsc{gadget}3 (update to \textsc{gadget}2, which was last described in \citealt{springel2005}). \eagle adopts cosmological parameters consistent with \citet{planck2014}, which are $\mathrm{\Omega_{m}}$ = 0.307, $\mathrm{\Omega_{\Lambda}}$ = 0.693, and $H_{0}$ = 67.77 km s\textsuperscript{-1} Mpc\textsuperscript{-1}. We use the reference model \textsc{Ref-L100N1504} with a co-moving volume of 100\textsuperscript{3} cMpc\textsuperscript{3} and the initial total number of particles is 2 $\times$ 1504\textsuperscript{3} (dark matter and baryons combined). The reference model is run with initial gas and dark matter particle masses of $m_{\rm gas}$ = 1.8 $\times$ 10\textsuperscript{6} \Msun and $m_{\rm DM}$ = 9.7 $\times$ 10\textsuperscript{6} \Msun, respectively. The mass of a stellar particle is roughly equivalent to that of the gas particle mass at its formation. We highlight some of the subgrid recipes relevant to the results here and direct the reader to \citet{schaye2015} and \citet{crain2015} for more details otherwise. Stars form in a stochastic manner where the density of the gas exceeds a minimum threshold that is metallicity-dependent \citep{sd2008}, following a \citet{chabrier2003} IMF. Energy from both stellar and AGN feedback are injected thermally and stochastically into the neighbouring particles. Thus, AGN feedback operates in a single heating mode. The free parameters of the simulation were fine-tuned to match the observed galaxy stellar mass function, galaxy sizes, and the stellar-to-BH mass at $z \sim 0$. 

\subsection{\magc}
\label{subsec:magc}
The \magc (\textit{Magneticum Pathfinder}; \citealt{teklu2015, dolag2017, schulze2018}) simulations are a suite of cosmological hydrodynamical simulations of varying box sizes and resolutions, built on the $N$-body/SPH code \textsc{gadget}3. \magc adopts the Wilkinson Microwave Anisotropy Probe 7 (WMAP7; \citealt{komatsu2011}) cosmology, for which the parameters are the following: $\mathrm{\Omega_{m}}$ = 0.272, $\mathrm{\Omega_{\Lambda}}$ = 0.728, and $H_{0}$ = 70.4 km s\textsuperscript{-1} Mpc\textsuperscript{-1}. We use the \textsc{box4-uhr} run, which has a co-moving volume of 68\textsuperscript{3} cMpc\textsuperscript{3} and an initial total number of particles of 2 $\times$ 576\textsuperscript{3} (gas and dark matter combined). The initial gas and dark matter particle masses are 1.0 $\times$ 10\textsuperscript{7} and 5.1 $\times$ 10\textsuperscript{7} \Msun, respectively. Each gas particle can spawn up to 4 stellar particles, such that the average mass of a stellar particle is 1.8 $\times$ 10\textsuperscript{6} \Msun. SF and stellar feedback are both based on the model introduced in \citet{sh2003}, where SF follows a \citet{chabrier2003} IMF. The prescription for AGN feedback follows \citet{fabjan2010}, such that BHs can switch between quasar and radio mode feedback depending on whether the accretion rates exceed a given Eddington ratio or not. Both modes operate in the form of thermal feedback. The radio mode is designed to have a 4 times larger feedback efficiency to account for massive BHs having low accretion rates but efficient in producing mechanical energy that heats the surrounding environment. Unlike \eagle and \tng, \magc is calibrated to reproduce the hot gas embedded in the intracluster medium of observed galaxy clusters, but not fine-tuned to match the galaxy stellar mass function at $z \sim 0$. 

\magc has the lowest particle resolution and smallest box size out of the three simulations, where on a galaxy scale, we can only extract robust measurements for galaxies at \Mstar $>$ 5 $\times$ 10\textsuperscript{9} \Msun. We take this difference in stellar mass threshold into account when comparing results with MAGPI and other simulations, as shown later in Section \ref{sec:res}. 

\subsection{\tng}
\label{subsec:tng}
The \tng project (The Next Generation Illustris; \citealt{naiman2018, springel2018, marinacci2018, nelson2018, pillepich2018b}) consists of a suite of cosmological magnetohydrodynamical simulations that directly build upon the physical framework of the \textsc{Illustris} \citep{genel2014, vogelsberger2014, sijacki2015} simulations. Like its predecessor, \tng is run on \textsc{arepo} \citep{springel2010}, which employs Moving Voronoi Mesh (MVM) for magnetohydrodynamics and TreePM for gravity. \tng adopts the cosmological parameters from \citet{planck2016}, with $\mathrm{\Omega_{m}}$ = 0.3089, $\mathrm{\Omega_{\Lambda}}$ = 0.6911, and $H_{0}$ = 67.74 km s\textsuperscript{-1} Mpc\textsuperscript{-1}. We use the TNG100-1 run for our analysis, which has a co-moving volume of $\approx$111\textsuperscript{3} cMpc\textsuperscript{3} and 1820\textsuperscript{3} particles for dark matter. We refer the reader to any of the introductory papers mentioned previously, or \citet{weinberger2017} and \citet{pillepich2018a}, for more details on the galaxy formation model and numerics, but we highlight some relevant processes here. Like \eagle and \magc, star formation is implemented stochastically with a \citet{chabrier2003} IMF, but with a threshold that is only set by density, not metallicity \citep{sh2003}. Supernova-driven stellar feedback is directly launched from the star-forming gas as outflows, where the energy injected via winds consists of both kinetic and thermal components. AGN feedback is implemented in dual mode, where at low accretion rates relative to the Eddington limit, kinetic energy is injected into the surroundings in the form of BH-driven winds, whereas thermal feedback is invoked at higher accretion rates. The kinetic feedback mode is known to be responsible for quenching galaxies above a BH mass threshold of 10\textsuperscript{8.2} \Msun (or \Mstar threshold of 10\textsuperscript{10} \Msun; \citealt{terrazas2020, piotrowska2022}) at $z \sim 0$. The \tng simulations are calibrated to match the following observables at $z \sim 0$: the galaxy stellar mass function, stellar-to-halo mass relation, BH mass to galaxy or halo mass relation, halo gas fraction, and galaxy stellar sizes. 

\section{Methods}
\label{sec:methods}

\subsection{\simspin outputs}
\label{subsec:simspin}
Galaxies have been selected from each simulation using their respective halo finder catalogues. \eagle, \tng, and \magc all use the \textsc{subfind} halo finder algorithm \citep{springel2001, dolag2009} though we note the detailed implementation of this code may vary from project to project. We consider only subhalos with associated stellar masses above 10\textsuperscript{9} \Msun (or 5 $\times$ 10\textsuperscript{9} in \magc), as defined by their respective algorithms. For each identified subhalo centre of potential, we select all associated sub halo particle flavours within a 50 kpc (physical) aperture for processing with \simspin. This choice allows us to remove erroneously identified ongoing mergers from our selection of galaxies.

Mock data cubes consist of two spatial directions in projection ($xy$) and kinematic information along the line of sight ($z$). These are generated to match MAGPI observations, including the average PSF full-width half-maximum (FWHM) of 0.6 arcsec (based in the $r$-band), the average line spread function (LSF) FWHM of 2.51 \AA, and the pixel scale of 0.2 arcsec pixel\textsuperscript{-1}.  The instantaneous\footnote{Independent of the time step associated with the snapshot.} SFR and stellar mass measurements are taken directly from the simulations, where a measurement is tagged to each particle/cell. We apply the particle/cell positions to a grid at the MUSE spatial resolution. FITS file outputs of SFR and stellar mass maps of each galaxy are then extracted by collapsing the cubes along the line of sight. 

For the stellar particles, each is assigned a luminosity using spectra from the \textsc{e-miles} \citep{vazdekis2016} templates. Given a particle’s age, metallicity and initial mass, as returned by all three simulations, we have interpolated the \textsc{e-miles} grid and convolved resulting spectral templates with the SDSS $r$-band filter to obtain an observed $r$-band flux for each system in projection using consistent methodology. The mock $r$-band maps are later used to define the galaxy centre needed to measure radial profiles (as described in Section \ref{subsec:rprof}). We neglect contributions from dust emission in these mock spectra, as none of the simulations considered trace the evolution of dust. 

The \simspin outputs do not have observational noise (e.g., signal-to-noise, CCD read-out noise, dithering patterns) added. But due to finite particle sampling of mass distributions, there is some effect due to shot noise within the simulations \citep{harborne2024}. 

\begin{figure}
    \centering
    \includegraphics[width=\columnwidth]{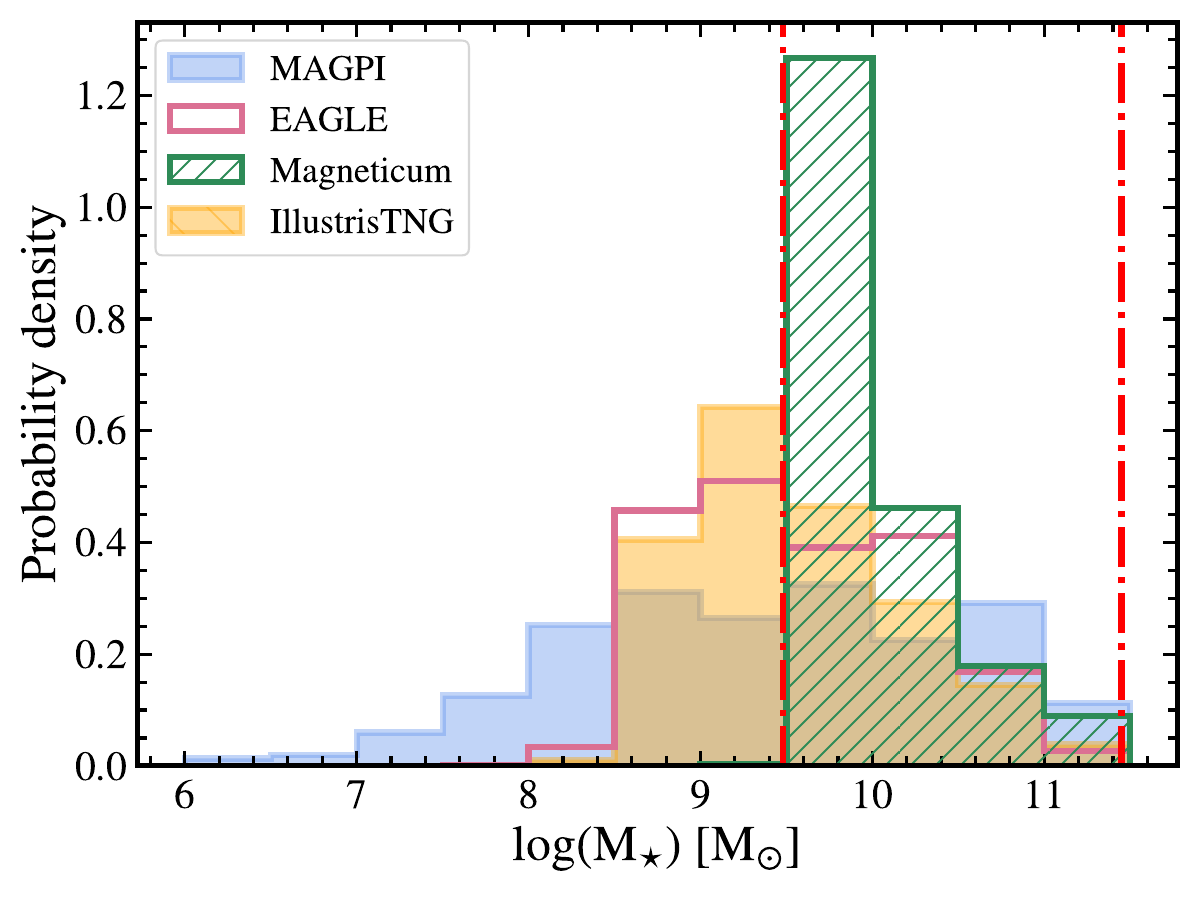}
    \caption{Histograms of the measured \Mstar for the MAGPI sample and each simulation. The MAGPI sample probes down to much lower masses, whereas the simulations can only probe down to 10\textsuperscript{8} \Msun, mostly due to numerical resolution limits. The red dash-dotted lines indicate the range of \Mstar we measure radial trends for, where the lower and upper cut offs are set by \magc and MAGPI, respectively.}
    \label{fig:mstar_hist}
\end{figure}

\begin{figure}
    \centering
    \includegraphics[width=\columnwidth]{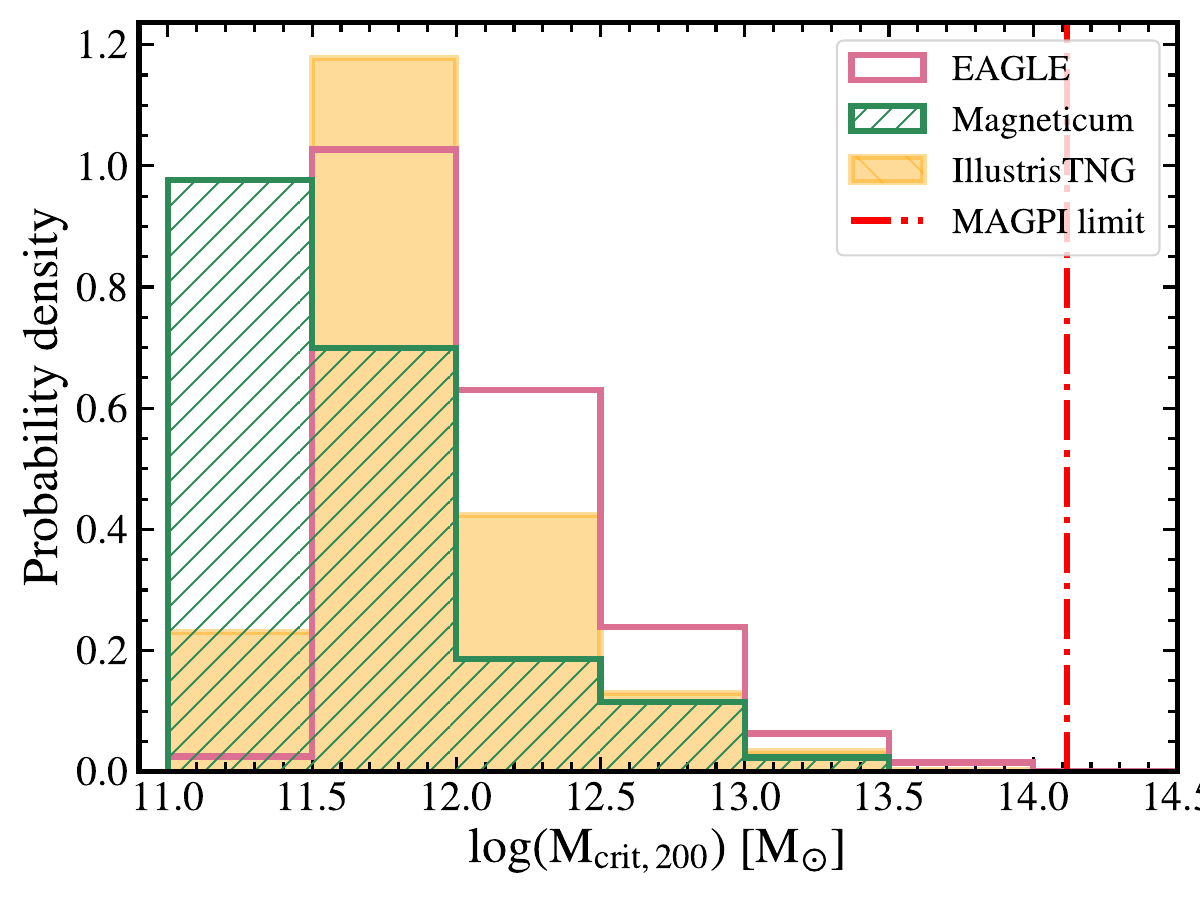}
    \caption{Histograms of the measured \Mcrit for central galaxies with 9.7 $\leq$ $\log_{10}$(\Mstar/\Msun) $\leq$ 11.4 (i.e., stellar masses in the range indicated by the red dash-dotted lines in Fig. \ref{fig:mstar_hist}) in each simulation, following the same colour scheme as Fig. \ref{fig:mstar_hist}. The \Mcrit distributions are comparable between \eagle and \tng, whereas a larger fraction of centrals reside in lower mass halos in \magc. The red dash-dotted line indicates the upper limit in \Mcrit set by the MAGPI sample.}
    \label{fig:mc200comp}
\end{figure}

\subsection{Galaxy selection criteria}
\label{subsec:sample}
We run \simspin on all galaxies with star-forming (i.e., non-zero SFRs) gas particles/cells found in the corresponding $z \sim 0.3$ snapshot of each simulation. There are a total of 8558 such galaxies in \eagle, 1383 in \magc, and 17769 in \tng. We place `detection' limits on the observable SFRs and stellar masses on a spaxel-by-spaxel basis, which leads to a cut on the number of galaxies. To determine the limit for SFRs, we take the lowest measured SFR of MAGPI galaxies on a spaxel-by-spaxel basis. In \citetalias{mun2024}, we used two SFR indicators -- \Halpha and D4000 -- to probe the star-forming and passive populations in MAGPI. For spaxels classified as star-forming by the Baldwin, Phillips \& Terlevich (BPT; \citealt*{bpt1981}) diagram, we measured dust-corrected \Halpha-based SFRs using the Balmer decrement. For all other spaxels (i.e., either classified as composite/AGN or non-detections), we measured the D4000-based SFRs from the D4000-sSFR relation. We refer the reader to Section 3.3 in \citetalias{mun2024} for more details on the methodology. The lower limit for all \Halpha and D4000-based SFRs is measured to be $\log_{10}$(SFR/\Msun yr\textsuperscript{-1}) $\approx$ -4.35 (or $\log_{10}$[\sigsfr/\sigsfrunits] $\approx$ -4.27), which is then applied on a spaxel-by-spaxel basis to the star-forming gas maps from each simulation. 

We also place limits on the observable stellar mass surface density (i.e., \sigstar) to ensure the measurements do not suffer from poor particle sampling. For the \eagle simulations, \citet{trayfordschaye2019} excluded pixels with less than 5 stellar particles. We examine the distribution of spaxels in SFR and stellar mass surface density (\sigsfr--\sigstar) space in all three simulations to catch any obvious signs of the measurements hitting the mass resolution limit (i.e., artificial stripes of spaxels with identical \sigstar values). \magc has the lowest resolution, where we found the limit in \sigstar to be at $\log_{10}$(\sigstar/\sigstarunits) $\approx$ 7.1 (or $\log_{10}$[\Mstar/\Msun] $\approx$ 7), which roughly corresponds to 5 stellar particles per spaxel. We implement this cut in stellar mass, again on a spaxel-by-spaxel basis, to the stellar mass maps across all simulations. 

With the final sample, we then measure total SFRs and \Mstar by summing all spaxels that pass the SFR and \Mstar cut for each galaxy. Applying these limits result in a small sample of galaxies with measured total SFRs and stellar masses of 0, where a total of 12 galaxies from \eagle, 1 galaxy from \magc, and 48 galaxies from \tng were removed. The final sample sizes for each simulation are reported in Table \ref{tab:sims_info}. We show the distribution of total stellar masses for MAGPI and each simulation in Fig. \ref{fig:mstar_hist}. As mentioned earlier in Section \ref{subsec:magc}, we implement a cutoff at 5 $\times$ 10\textsuperscript{9} \Msun in \magc, such that we are probing a relatively narrow distribution of \Mstar for \magc. 

\begin{figure*}
    \centering
    \begin{subfigure}{\textwidth}
        \includegraphics[width=\linewidth]{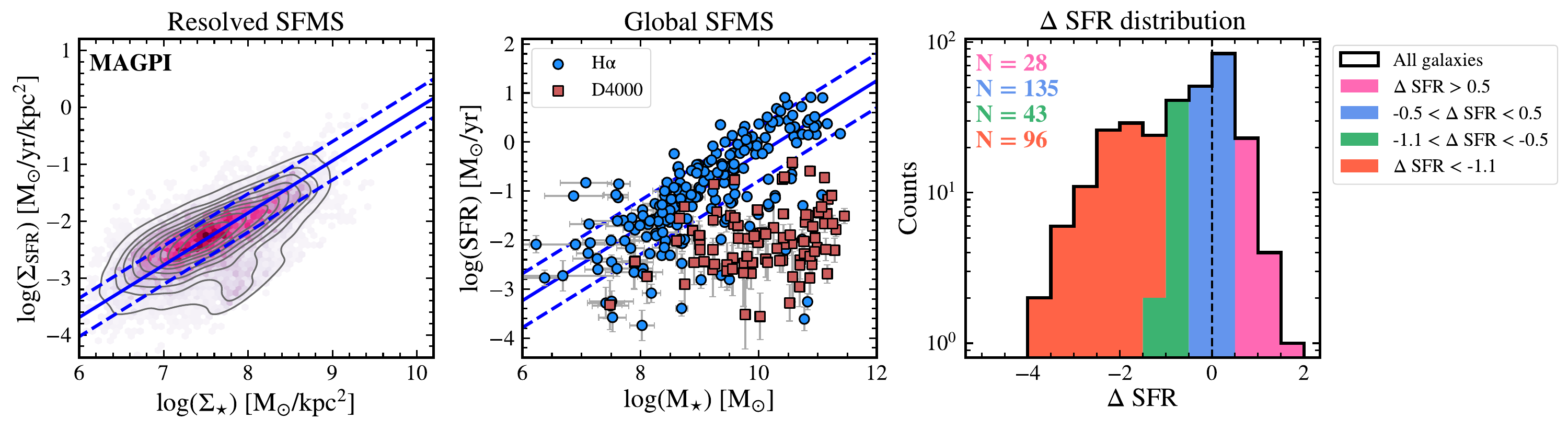}
    \end{subfigure}
    \begin{subfigure}{\textwidth}
        \includegraphics[width=\linewidth]{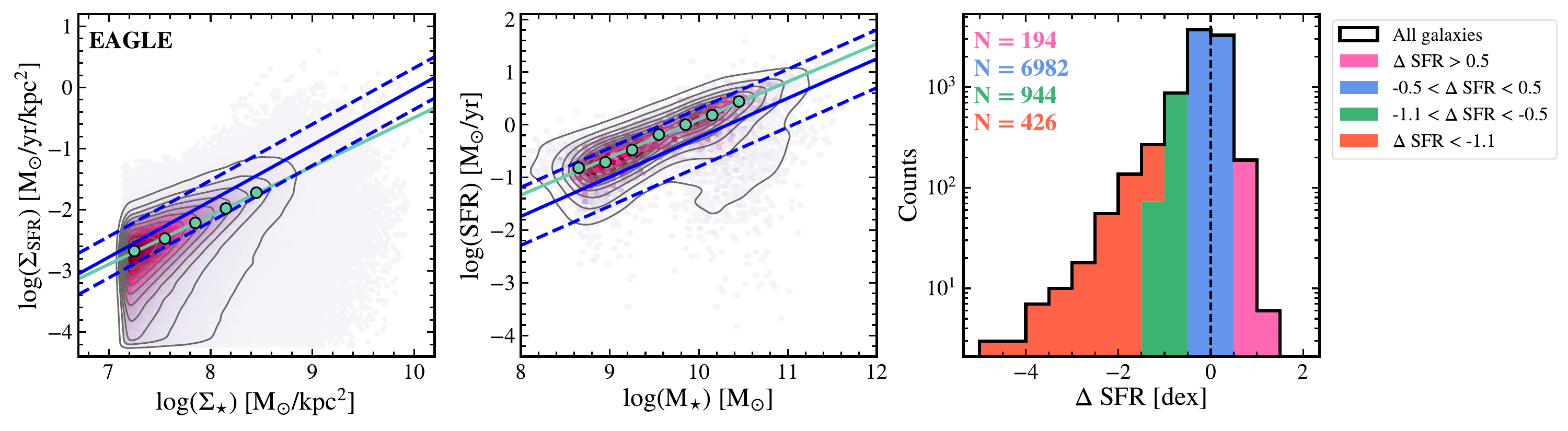}
    \end{subfigure}
    \begin{subfigure}{\textwidth}
        \includegraphics[width=\linewidth]{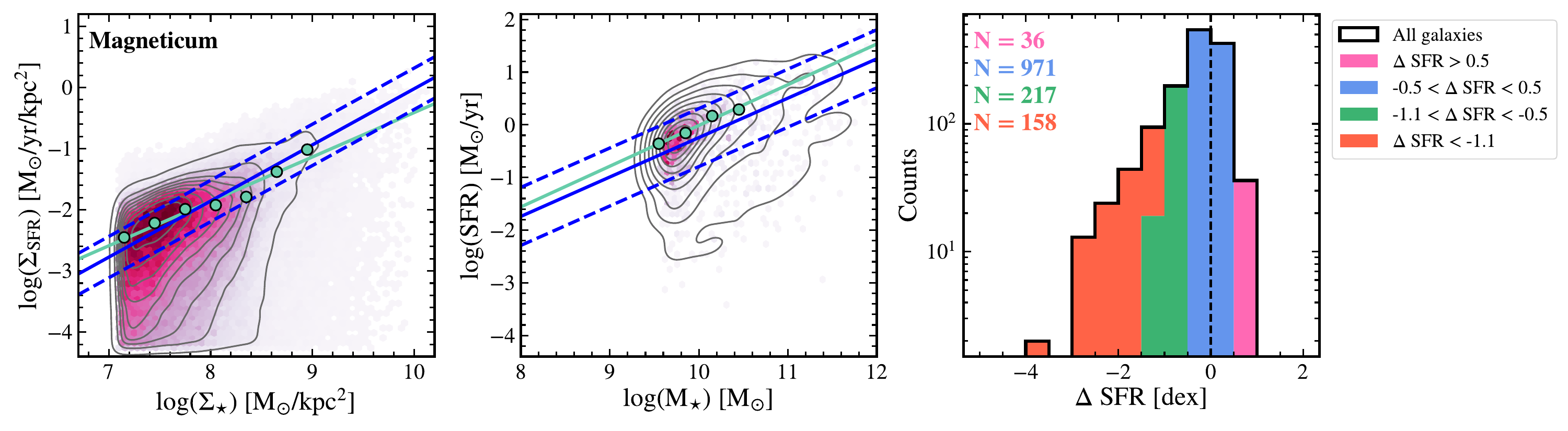}
    \end{subfigure}
    \begin{subfigure}{\textwidth}
        \includegraphics[width=\linewidth]{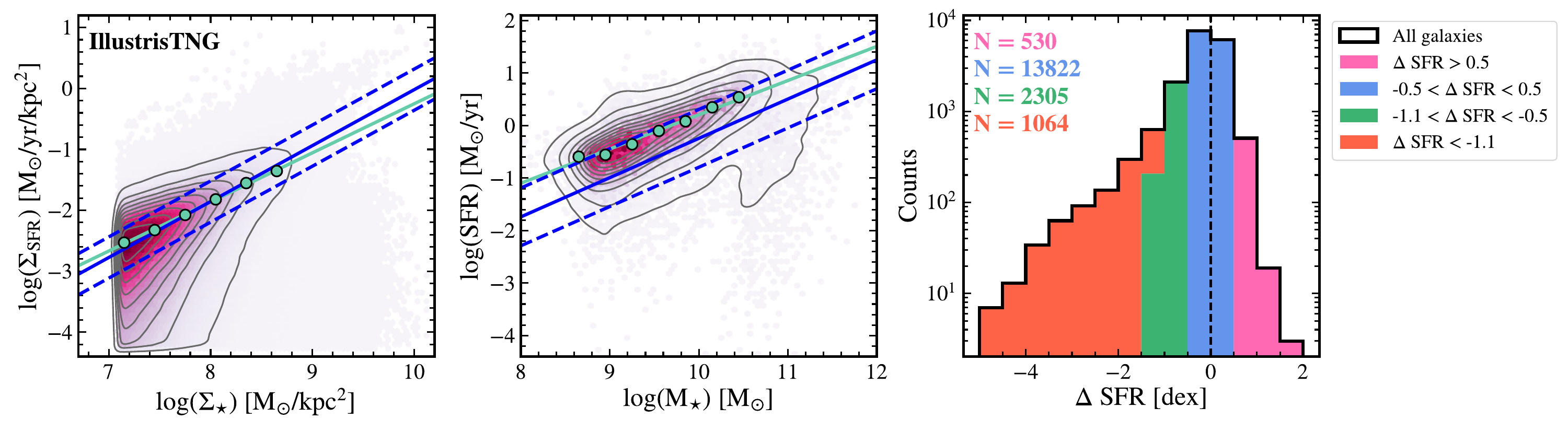}
    \end{subfigure}
    \caption{Top to bottom: resolved (left column) and global (middle column) SFMS fits and \dsfr histograms (right column) for MAGPI, \eagle, \magc, and \tng. Both the resolved and global SFMS fits for MAGPI are measured based on \Halpha-detected SF spaxels/galaxies. The best fit global and resolved SFMS for MAGPI with the root mean square (RMS) error are overlaid as blue solid and dashed lines, respectively, for each simulation as reference. For MAGPI, we show the full distribution of galaxies from \citetalias{mun2024}, where blue circles denote \Halpha detections and red squares D4000 detections. For the simulations, the resolved and global SFMS fits are based on the peak of the \sigsfr or SFR PDF of each \sigstar or \Mstar bin (see Section \ref{subsec:metrics}), which are denoted by the turquoise data points. The MAGPI sample probes down to lower values of $\log$(\sigstar) and $\log$(\Mstar), such that the horizontal axis ranges between MAGPI and the simulations are not equal. With the exception of the normalisation, the distribution of each simulation data agrees well with the slope of the MAGPI global SFMS. In a resolved sense, MAGPI measures a steeper resolved SFMS, although a broadly good agreement is seen in the normalisations. Similarities are also observed in the overall distribution of \dsfr, although \eagle and \tng both consist of a notable population of passive galaxies with low \dsfr.}
    \label{fig:sfmsdsfrsims}
\end{figure*}

\begin{table*}
    \centering
    \caption{Coefficients of the global and resolved SFMS fits for MAGPI observations (\citetalias{mun2024}) and simulations used in this study.}
    \begin{tabular}{ccccc}
    \hline
    Observation/Simulation & SFMS slope & SFMS intercept & Resolved SFMS slope & Resolved SFMS intercept \\
    \hline
    MAGPI & 0.75 $\pm$ 0.04 & -7.73 $\pm$ 0.06 & 0.92 $\pm$ 0.01 & -9.20 $\pm$ 0.01 \\
    \eagle & 0.72 $\pm$ 0.03  & -7.07 $\pm$ 0.29 & 0.80 $\pm$ 0.02 & -8.49 $\pm$ 0.14 \\
    \magc & 0.77 $\pm$ 0.08 & -7.75 $\pm$ 0.83 & 0.73 $\pm$ 0.07 & -7.67 $\pm$ 0.60 \\ 
    \tng & 0.65 $\pm$ 0.05 & -6.31 $\pm$ 0.42 & 0.81 $\pm$ 0.02 & -8.30 $\pm$ 0.14 \\ 
    \hline
    \end{tabular}
    \label{tab:sfms_fits}
\end{table*}

\subsection{Environmental parameters}
\label{subsec:suppdata}
To complement our analysis of the environmental effects on radial profiles, we use the supplementary catalogs available for each simulation. Specifically, we use the central/satellite classifications and \Mcrit, the latter being the mass of the host halo enclosed within a sphere of an average density that is 200 times the critical density of the universe. The flagging of centrals/satellites is done similarly across all simulations, based on friends-of-friends (FoF) halos found with \textsc{subfind} \citep{springel2001, dolag2009}. Generally, central galaxies are flagged as the most massive galaxy in the FoF halo; any other subhalo is a satellite. The total number of centrals and satellites used in the final sample for each simulation are reported in Table \ref{tab:sims_info}. We show a set of histograms of the \Mcrit distributions for each simulation in Fig. \ref{fig:mc200comp}. 

For the MAGPI sample, our classification of central/satellite is roughly based off of `primary'/`secondary' status, where `primary' galaxies were purposely selected to be GAMA galaxies measured as the central galaxy of their local environments; all other MAGPI galaxies are `secondary' galaxies found within the FOV and redshift of the `primary' and are thus considered as satellites. We cross-check these classifications with an FoF algorithm (following the methodology outlined in \citealt{knobel2009} and \citealt{robotham2011}, as used for the GAMA survey) run on all MAGPI observed galaxies within the MUSE FOV (Harborne et al. in preparation). We find that all `primary' galaxies are consistent with being the brightest central galaxy (as according to their $i$-band luminosities) in their respective group environments. 

We also use preliminary measurements of dynamical group masses via the same FoF algorithm to measure the maximum halo mass of MAGPI galaxies. While these masses have been fine-tuned to the Deep Extragalactic VIsible Legacy Survey (DEVILS; \citealt{davies2018}; Bravo et al. in preparation), a deep spectroscopic survey of $\sim$60000 galaxies down to $Y < 20.68$ mag (90 per cent completeness limit) over $\sim$6 deg\textsuperscript{2}, the algorithm has only been tested on MAGPI galaxies within the MUSE FOV. As potential group members outside the FOV have not been accounted for, we only use these group masses to measure an upper limit on \Mcrit, i.e., $\log$(\Mcrit/\Msun) = 14.1, to be implemented on the simulations as well. This limit is indicated in Fig. \ref{fig:mc200comp} as the red dash-dotted line. Given that larger simulation box sizes ($>$100 cMpc) are needed to probe a significant number of dense environments with $\log$(\Mcrit/\Msun) $>$ 14, this cut off removes a small portion of the sample from each simulation used here: 89 ($\sim$1 per cent) galaxies from \eagle, 6 ($\sim$0.4 per cent) from \magc, and 176 ($\sim$1 per cent) from \tng. 

\subsection{Global and local star formation metrics}
\label{subsec:metrics}
We adopt the same global and local SF metrics used in \citetalias{mun2024} for comparison purposes. Similar in definition to the commonly used sSFR, \dsfr and \dssfr measure the logarithmic offset from the global and resolved SFMS, respectively \citep{ellison2018, bluck2020}. The two parameters are defined below: 

\begin{align} 
    \Delta \mathrm{SFR} &= \log_{10}(\mathrm{SFR_{gal}}) - \log_{10}(\mathrm{SFR_{MS}}), \label{eq:dsfr} \\
    \Delta \mathrm{\Sigma_{SFR}} &= \log_{10}(\mathrm{\Sigma_{SFR, spax}}) - \log_{10}(\mathrm{\Sigma_{SFR, MS}}). \label{eq:dssfr} 
\end{align}

The parameters are defined such that positive $\Delta$-values indicate higher SF activity on a galaxy/spaxel level with respect to the global/resolved SFMS, and vice versa. We then use the \dsfr parameter to classify galaxies in each simulation into different star-forming states, using the following definitions: 
\begin{enumerate}
    \item above the SFMS: \dsfrsb dex,
    \item SFMS: -0.5 dex $<$ \dsfr $<$ 0.5 dex,
    \item just below the SFMS: -1.1 dex $<$ \dsfr $<$ -0.5 dex,
    \item far below the SFMS: \dsfrq dex,
\end{enumerate}
where these definitions were taken from \citet{bluck2020} and adopted in \citetalias{mun2024} as well for a fair comparison to studies based on the local Universe. 

Given the definition of the two parameters in Equations \ref{eq:dsfr} and \ref{eq:dssfr}, we require robust definitions of the SFMS. In order to ensure that the adopted SFMS represents the distribution of galaxies well, we measure the SFMS separately for each simulation. For the MAGPI SFMS, we adopt the same definition in \citetalias{mun2024}, for which both the global and resolved SFMS were derived based on the sum and distribution of \Halpha-detected spaxels classified as star-forming by BPT diagram, respectively. We assume the SFMS follows a single power law, translating into a linear relation in log-log space. Given that in simulations, we are unable to distinguish between star-forming and passive populations of galaxies/spaxels as commonly done in observations using \Halpha detections, we follow a different approach here. The general process is the following: we first measure the probability density function (PDF) of SFR (or \sigsfr) in bins of \Mstar (or \sigstar), using Gaussian kernels of smoothing bandwidths measured using Silverman's rule with \textsc{scipy} \citep{virtanen2020}. We then use the peak of the PDF to fit the SFMS with orthogonal distance regression (ODR). The values and errors for the resolved and global SFMS fits are shown in Table \ref{tab:sfms_fits}. We show the resulting resolved and global SFMS fits for each simulation, along with the MAGPI fits for comparison purposes, in Fig. \ref{fig:sfmsdsfrsims}. The slopes of the global SFMS agree within 1$\sigma$ for the three simulations. The slopes of the global SFMS measured for \eagle and \magc are within 1$\sigma$ agreement with that of the MAGPI SFMS, whereas \tng shows agreement within 2$\sigma$. On the other hand, both the slopes and intercepts of the resolved SFMS of all three simulations are shallower than the MAGPI SFMS. The slopes across the three simulations agree with each other within 1$\sigma$. The \dsfr distribution for each sample is shown in the right column of the figure, where both \tng and \eagle show a notable fraction of passive galaxies, reaching below \dsfr = -4 dex. 

To fit the SFMS, we use galaxies in \eagle and \tng that can have as few stellar particles as 100, in contrast to \magc, where we only use galaxies with \Mstar $>$ 5 $\times$ 10\textsuperscript{9} \Msun, corresponding roughly to 2800 stellar particles. The former selection criteria has been shown to be sufficient to get converged results for the slope and zero point of the SFMS in both \eagle and \tng simulations \citep[e.g.,][]{furlong2015, donnari2019} and even to measure the scatter of the SFMS \citep{mattheeschaye2019, katsianis2019}. The reason for this is that convergence in global measurements of stellar mass and SFR is much easier to achieve than convergence in galaxy sizes \citep{ludlow2019, ludlow2020} and kinematics \citep{harborne2020, harborne2023}. In this paper, we thus use the lower mass sample to fit the main sequence, but then apply more stringent stellar mass constraints to study the \dssfr radial profiles, as explained in Section \ref{sec:res}. 

\subsection{Radial profiles}
\label{subsec:rprof}
We measure radial profiles of \dssfr for each global SF state as defined in Section \ref{subsec:metrics}, as done in \citetalias{mun2024}. To account for inclination, we measure the profiles in bins of elliptical annuli, where we use \textsc{ProFound}\footnote{\url{https://github.com/asgr/ProFound}} \citep{robotham2018} to measure structural parameters such as effective radii (i.e., \re), position angles, and axial ratios. \textsc{ProFound} is run directly on the \simspin outputs. The parameters given by \textsc{ProFound} are PSF-convolved measurements, where the size and shape of the PSF is not accounted for in source detection. 

We calculate galactocentric distances normalised by \re on a spaxel-by-spaxel basis, where the galaxy centre is defined as the location of the peak stellar flux. We then radially bin spaxels with measured \dssfr into bins of width 0.5 \re from 0 to 5 \re, where we take the median of each bin to build a population-averaged radial profile for each global SF state. All radial bins with at least 10 galaxies ($\mathrm{N_{gal}} \geq$ 10) and 50 spaxels ($\mathrm{N_{spax}} \geq$ 50) are included in the generation of the profiles. We measure the error on the median by bootstrap, where we randomly sample 100 distributions of \dssfr for each bin with replacement, and then take the standard deviation on the medians of the samples as the error. We note that these errors indicate the standard error on the median, and as such may not represent the overall scatter (on average $\sim \pm$0.4 dex) in the individual galaxy profiles. We show the profiles with the 25th and 75th percentiles for all three simulations, along with MAGPI, in Fig. \ref{fig:rprof_percent} in Appendix \ref{sec:app3}. 

We note that, for this work, we explore the properties of our simulated galaxies farther out than in \citetalias{mun2024}. This was done to ensure we capture any environmental effects only observable in the far outskirts. In contrast, our MAGPI observations are limited to $\sim$3 \re, due to additional selections performed in our observations, such as multiple emission line S/N cuts to identify SF regions and correct for dust extinction for the use of \Halpha as a SFR indicator, and uncertainties in the stellar mass measurements (see Sections 3.2 and 3.3 in \citetalias{mun2024}). The simulations also benefit from higher number statistics (see Table \ref{tab:sims_info} and Fig. \ref{fig:sfmsdsfrsims}) relative to MAGPI observations. Nevertheless, this works as an advantage of using simulations, where we can potentially capture the effects of environmental mechanisms more readily. 

\begin{figure*}
    \centering
    \includegraphics[width=\linewidth]{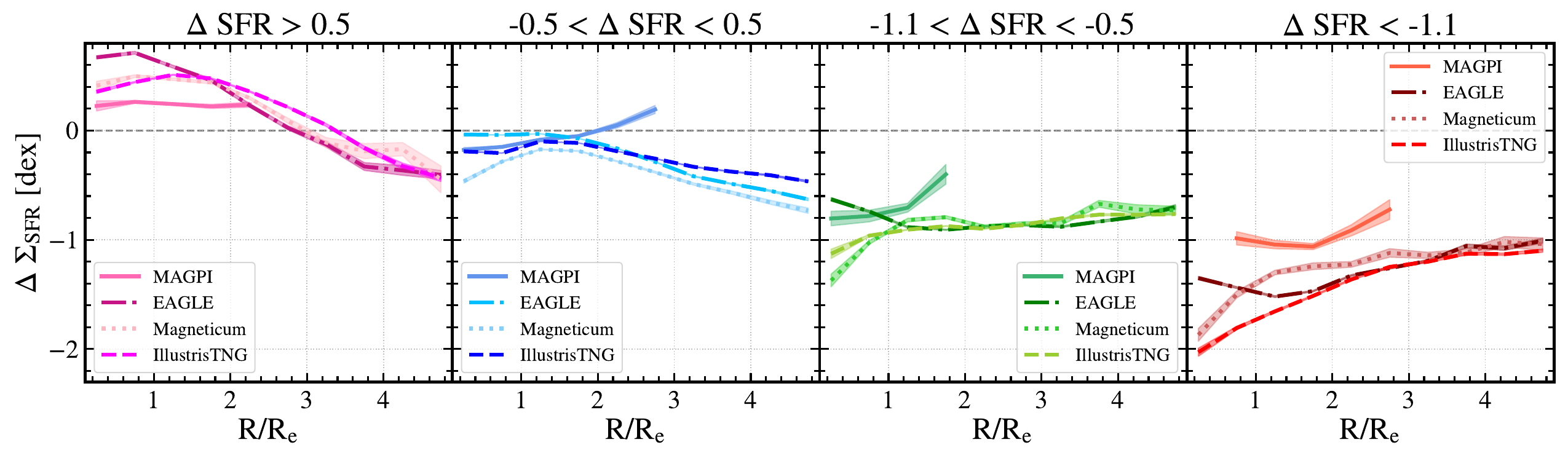}
    \caption{Median \dssfr profiles as observed in the respective $z \sim 0.3$ snapshots for \eagle (dash-dotted), \magc (dotted), and \tng (dashed), plotted along with the MAGPI (solid) sample. The profiles are measured with respect to the corresponding resolved SFMS for each sample. Only galaxies with a) 9.7 $\leq$ $\log_{10}$(\Mstar/\Msun) $\leq$ 11.4, and b) $\log_{10}$(\Mcrit/\Msun) $\leq$ 14.1 are shown for all samples. The panels show the profiles for each global SF state in decreasing order of \dsfr from left to right. The grey dashed line of \dssfr = 0 is shown as reference to show where the profiles fall with respect to the resolved SFMS. The shaded regions indicate the bootstrap errors based on the median. While there is a broad agreement in the overall trends among the simulations, notable discrepancies are captured particularly within 1 \re, where differing degrees of central suppression are captured, specifically for galaxies on and below the SFMS. The trends observed in the simulations are also at odds with those of the MAGPI sample. We discuss the implications further in Section \ref{subsec:magpi_comp} and \ref{sec:disc}.}
    \label{fig:allsimrprof}
\end{figure*}

\section{Results}
\label{sec:res}
In this section, we present the \dssfr radial profiles for galaxies from the \eagle, \magc, and \tng simulations and MAGPI observations at $z\sim0.3$. We show the profiles in Fig. \ref{fig:allsimrprof}, where we implement stellar (9.7 $\leq$ $\log_{10}$[\Mstar/\Msun] $\leq$ 11.4) and halo ($\log_{10}$[\Mcrit/\Msun] $\leq$ 14.1) mass cuts set by \magc and MAGPI (see Fig. \ref{fig:mstar_hist} and \ref{fig:mc200comp}), for a fairer comparison across all samples where possible. The lower and upper limits of the stellar mass cuts are set by the mass resolution of \magc and the maximum \Mstar of MAGPI, respectively. The upper limit of halo mass is set by the maximum \Mcrit of MAGPI. This results in a final sample of 3539 ($\sim$41 per cent of the original sample) galaxies from \eagle, 1351 ($\sim$98 per cent) galaxies from \magc, and 6434 ($\sim$36 per cent) from \tng.

We begin by discussing the radial trends from the simulations in Section \ref{subsec:rprof_sims}, exploring the implications of how SF is regulated and quenched in galaxies at $z \sim 0.3$ from a theoretical perspective. We then include MAGPI observations in the discussion to complement the picture set forward by the simulations in Section \ref{subsec:magpi_comp}. For a fair comparison to all simulations, where the lowest stellar mass cut off is \Mstar $>$ 5 $\times$ 10\textsuperscript{9} \Msun (set by \magc), we re-generated radial profiles for MAGPI with this stellar mass cut off. This results in a new total of 181 ($\sim$60 per cent of the original 302) galaxies, which motivates a less stringent cut in $\mathrm{N_{gal}}$ and $\mathrm{N_{spax}}$ than the ones employed in \citetalias{mun2024}, where we now measure median \dssfr for all radial bins with $\mathrm{N_{gal}} \geq$ 5 and $\mathrm{N_{spax}} \geq$ 25. There is no significant difference in the MAGPI profiles when relaxing this criterion. We emphasize that this cut is only implemented on the MAGPI sample; all three simulations use the $\mathrm{N_{gal}} \geq$ 10 and $\mathrm{N_{spax}} \geq$ 50 even with the stellar and halo mass cuts, as explained in Section \ref{subsec:rprof}. 

\subsection{Radial trends in SF as seen by simulations} 
\label{subsec:rprof_sims}
We show the \dssfr profiles for galaxies at $z \sim$ 0.3 in \eagle, \magc, and \tng in Fig. \ref{fig:allsimrprof}. From left to right, we show the median profiles for each global SF state in decreasing order of \dsfr. Across all simulations, the profiles show broad general agreement moving from above to below the SFMS, with the median \dssfr profiles shifting to lower normalisations with the global SF state. On average, all three simulations show negative gradients in \dssfr for galaxies above and on the SFMS, where the former indicates centrally enhanced SF and the latter implies that galaxies on the SFMS may already be undergoing outside-in quenching. For galaxies below the SFMS, we begin to see positive gradients indicative of inside-out quenching. 

While the overall shapes of the profiles are broadly in agreement, we observe notable differences in the degree of SF suppression within $R \sim 1$ \re for galaxies on and below the SFMS. A central depression is seen in \magc across all star-forming states, except for galaxies above the SFMS. In contrast, the gradients within $R \sim 1$ \re from \eagle are negative for galaxies in the two \dsfr bins below the SFMS, whereas a central depression (i.e., positive gradient) is seen in \tng starting in the \dsfrgv (green) bin. The observed trends are likely due to a mixture of galaxies at different stages of evolution, i.e., some have yet to experience significant feedback. For \eagle, the lack of central SF suppression may be tied to previously observed positive gradients in stellar abundances of \eagle galaxies which manifest even in slow rotators \citep{lagos2022}. The findings reported in \citet{lagos2022} show that most \eagle galaxies at $z \sim 0$ have less $\upalpha$-enhanced centres associated with younger stellar ages. The latter is thought to be due to an excess of SF in the centre, which also leads to lower stellar velocity dispersions relative to the outskirts. This indicates that while AGN feedback in \eagle is effective at quenching galaxies globally (given the calibration to match the observed stellar mass function), that is not necessarily the case when studying the relative levels of SF in the centre vs. the outskirts. We discuss the role of AGN feedback on shaping the radial trends in each simulation further in Section \ref{subsec:disc_agn}. 

\begin{figure*}
    \centering
    \includegraphics[width=0.85\linewidth]{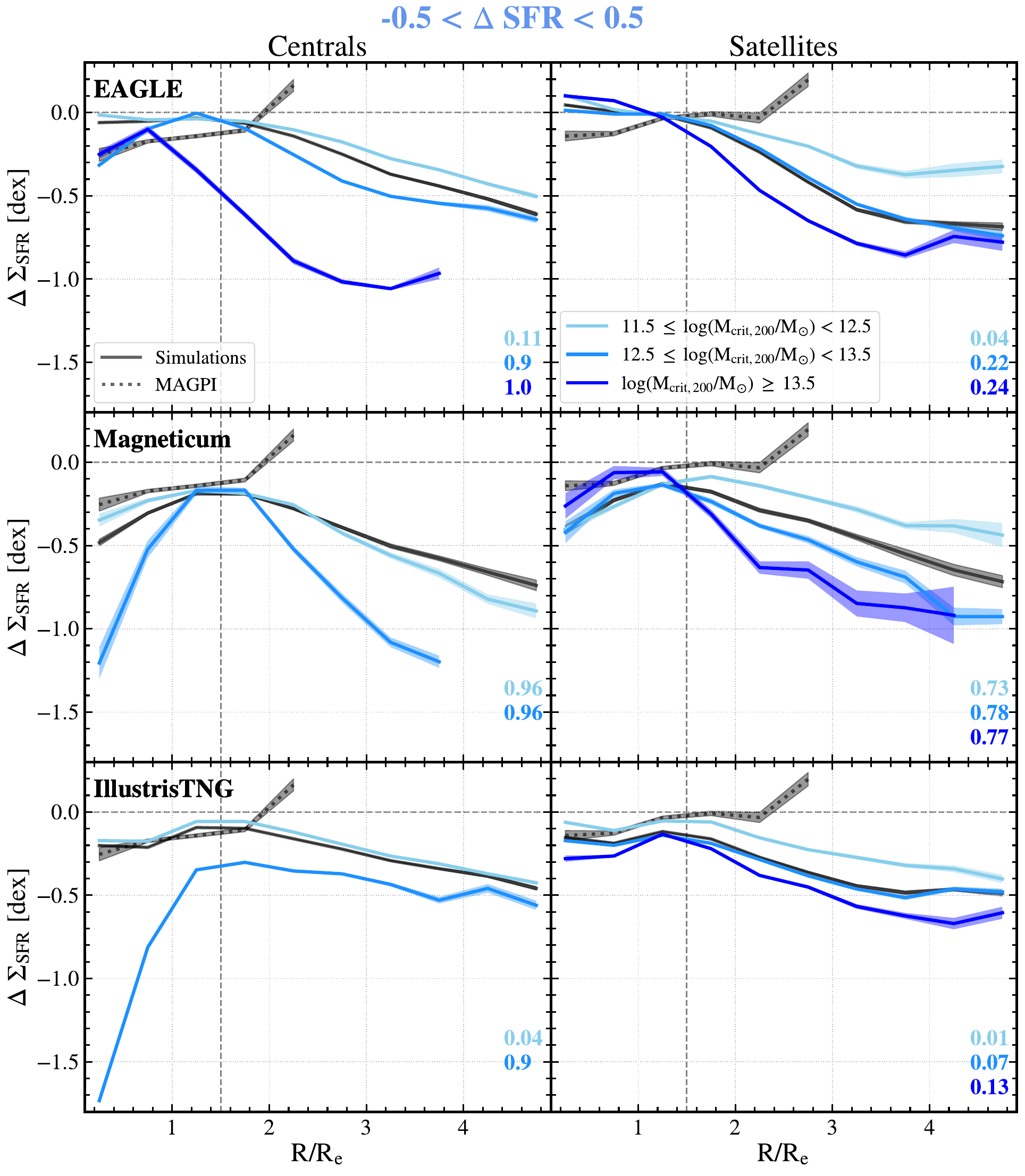}
    \caption{\dssfr profiles for SFMS galaxies (i.e., \dsfrms) shown separately for centrals (left column) and satellites (right column), for MAGPI (dotted) and all simulations (solid). The black lines in each panel show the median profiles for \textit{all} centrals and satellites, independent of \Mcrit. The solid lines in different shades of blue show the trends for different bins of \Mcrit for each simulation. The fraction of galaxies with stellar/BH masses above the known threshold that leads to quenching unique to each simulation, i.e. $\mathrm{f_{BH, THRESH}}$, is shown for each \Mcrit bin, following the same colour code as the profiles, in the bottom right corner of each panel. The grey vertical dashed line in each panel denotes $R = 1.5$ \re. A clear trend with environment is seen for all three simulations; centrals quench across all radii probed with stronger suppression in the centre, whereas satellites show increasingly suppressed SF particularly in the outskirts with increasing \Mcrit. However, this trend is only clear when controlling for both central/satellite status and \Mcrit, the profiles for centrals and satellites are otherwise indistinguishable for all four samples shown here.}
    \label{fig:ms_all_sims_comp}
\end{figure*}

We also note that we are averaging over a mixture of populations in Fig. \ref{fig:allsimrprof} -- centrals, satellites, and galaxies living across different environmental densities. As such, some notable trends or features may get washed out in these profiles. Previous studies have shown evidence for differences in dominant quenching mechanisms for central and satellite galaxies (e.g., \citealt{peng2012, davies2019, wright2019, bluck2020}). In particular, we expect central galaxies to be governed more so by internal processes such as AGN feedback, and satellites to be more vulnerable to external processes such as ram pressure stripping and tidal interactions. Furthermore, the degree of ram pressure stripping and frequency of interactions is known to vary with environment \citep{rh2005, brown2017, roberts2021, wang2022, khalid2024}, such that this motivates an investigation into how the observed radial trends vary with central/satellite status and halo mass. 

\begin{table*}
    \caption{Measured slopes for the inner ($R \leq 1.5$ \re) and outer ($R > 1.5$ \re) regimes of the radial profiles of central and satellite galaxies in the \dsfrms bin for MAGPI and all three simulations shown in Fig. \ref{fig:ms_all_sims_comp}.}
    \centering
    \begin{tabular}{ccccc}
    \hline
       \Mcrit bin & MAGPI & \eagle & \magc & \tng \\
         & ($R \leq 1.5$ \re, $R > 1.5$ \re) & ($R \leq 1.5$ \re, $R > 1.5$ \re) & ($R \leq 1.5$ \re, $R > 1.5$ \re) & ($R \leq 1.5$ \re, $R > 1.5$ \re) \\
        (\Msun) & (\slopeunits) & (\slopeunits) & (\slopeunits) & (\slopeunits) \\
    \hline
    \multicolumn{5}{c}{Centrals} \\
    \hline
       All & (0.11 $\pm$ 0.01, 0.54 $\pm$ 0.02) & (0.03 $\pm <$0.01, -0.18 $\pm <$0.01) & (0.29 $\pm$ 0.01, -0.18 $\pm <$0.01) & (0.11 $\pm <$0.01, -0.12 $\pm <$0.01) \\
       11.5 $\leq$ $\log_{10}$(\Mcrit) $<$ 12.5 & -- & (-0.02 $\pm <$0.01, -0.15 $\pm <$0.01) & (0.18 $\pm$ 0.02, -0.25 $\pm <$0.01) & (0.11 $\pm <$0.01, -0.12 $\pm$ 0.01) \\
       12.5 $\leq$ $\log_{10}$(\Mcrit) $<$ 13.5 & -- & (0.31 $\pm <$0.01, -0.17 $\pm <$0.01) & (1.04 $\pm$ 0.03, -0.52 $\pm$ 0.01) & (1.39 $\pm$ 0.01, -0.08 $\pm <$0.01) \\
       $\log_{10}$(\Mcrit) $\geq$ 13.5 & -- & (-0.09 $\pm$ 0.01, -0.17 $\pm$ 0.01) & -- & -- \\
    \hline
    \multicolumn{5}{c}{Satellites} \\
    \hline
       All & (0.10 $\pm$ 0.01, 0.2 $\pm$ 0.01) & (-0.06 $\pm <$0.01, -0.21 $\pm <$0.01) & (0.28 $\pm$ 0.01, -0.18 $\pm <$0.01) & (0.04 $\pm <$0.01, -0.11 $\pm$ 0.01) \\
       11.5 $\leq$ $\log_{10}$(\Mcrit) $<$ 12.5 & -- & (-0.12 $\pm <$0.01, -0.10 $\pm <$0.01) & (0.27 $\pm$ 0.01, -0.12 $\pm$ 0.01) & (0.01 $\pm <$0.01, -0.11 $\pm <$0.01) \\
       12.5 $\leq$ $\log_{10}$(\Mcrit) $<$ 13.5 & -- & (-0.02 $\pm <$0.01, -0.23 $\pm <$0.01) & (0.29 $\pm$ 0.02, -0.24 $\pm <$0.01) & (0.03 $\pm <$0.01, -0.10 $\pm <$0.01) \\
       $\log_{10}$(\Mcrit) $\geq$ 13.5 & -- & (-0.13 $\pm <$0.01, -0.18 $\pm <$0.01) & (0.2 $\pm$ 0.03, -0.23 $\pm$ 0.02) & (0.15 $\pm$ 0.01, -0.14 $\pm$ 0.01) \\
    \hline
    \end{tabular}
    \label{tab:ms_all_sims_slopes}
\end{table*}

In Fig. \ref{fig:ms_all_sims_comp}, we select the \dsfr bin with the highest sample statistics (i.e., SFMS galaxies; \dsfrms) and split the sample by central/satellite status into different bins of \Mcrit denoted by the different blue colours. We show the median \dssfr trends for \textit{all} centrals and satellites as black lines in the same panels. We show the full set of \dssfr profiles for the other global SF states for each simulation in Appendix \ref{sec:app1}, with the exception of the \dsfrsb bin, where low number statistics prevents a fair comparison of the trends between centrals and satellites. 

We examine the radial trends in three \Mcrit bins of width 1 dex, with a lower cut off at $\log_{10}$(\Mcrit/\Msun) = 11.5, for a fair comparison among all simulations (see Fig. \ref{fig:mc200comp}). We indicate the fraction of galaxies with stellar/BH masses above the known threshold for strong AGN feedback, $\mathrm{f_{BH, THRESH}}$, for each \Mcrit bin in each panel. 

In \eagle and \tng, AGN feedback is effective for galaxies with \Mstar $>$ 3 $\times$ 10\textsuperscript{10} \Msun \citep{bower2017} and $M_{\rm BH} \geq$ 10\textsuperscript{8.2} \Msun \citep{terrazas2020}, respectively. For \magc, there is no set mass threshold at which quenching via AGN feedback occurs; $\mathrm{f_{BH, THRESH}}$ simply indicates the fraction of galaxies with BHs, as not all satellites will have BHs seeded \citep{hirschmann2014}. We discuss these differences further in Section~\ref{sec:disc}.

We measure slopes of the profiles using linear fits with Monte Carlo derived uncertainties, where we fit `inner' ($R \leq 1.5$ \re) and `outer' ($R > 1.5$ \re) regions separately. The slopes and errors are shown in Table \ref{tab:ms_all_sims_slopes}. As observed in Fig. \ref{fig:ms_all_sims_comp}, the choice of 1.5 \re as a cut off is motivated by central depressions/elevations in \dssfr being mostly encapsulated within 1 -- 1.5 \re. We fit two radial regimes separately, rather than the full profile, to quantify the degree of internal and external processes occurring within the galaxies. The profiles are not always monotonic in these ranges, so the slopes are a coarse way to capture general trends. Nevertheless, the signs of the slopes indicate positive/negative gradients in SF, providing a simple diagnostic to compare the inner and outer regimes.  

We observe a clear trend with environment in \eagle, where centrals quench across all radii, with lower values of \dssfr towards more massive halos. In contrast, satellites show negligible changes in \dssfr within $R \sim$ 1.5 \re with increasing \Mcrit; it is only in the galaxy outskirts (most notably beyond $R \sim$ 2 \re) where we see increasing degree of suppression with increasing halo mass. The latter is not necessarily surprising, as here we are only focusing on galaxies with \dsfrms, so large deviations in \dssfr in the inner regions of galaxies would move galaxies away from the main sequence. The fraction of galaxies with \Mstar $>$ 3 $\times$ 10\textsuperscript{10} \Msun, i.e., $\mathrm{f_{BH, THRESH}}$, also increases with \Mcrit for both centrals and satellites, although the increase is more dramatic for centrals, reaching $\mathrm{f_{BH, THRESH}}$ = 1 in the highest halo mass bin. The trends above show that centrals and satellites undergo distinct quenching processes in \eagle; while both internal and external mechanisms govern radial trends in centrals, satellites seem to be governed only by environmental mechanisms acting in the outskirts. Dominant environmental processes in \eagle acting on satellites include ram pressure stripping and tidal interactions between galaxies \citep{marasco2016, bahe2017, wright2022, wang2023}.

Similar to \eagle, the centrals in \magc experience SF suppression at all radii. However, the central suppression is stronger in the inner profile at fixed halo mass. The profiles shift to lower normalisations with increasing \Mcrit for centrals, where a notably stronger degree of central suppression is captured at 12.5 $\leq \log_{10}$(\Mcrit/\Msun) $<$ 13.5 (0.18 to 1.04 \slopeunits). No profile is measured for the $\log_{10}$(\Mcrit/\Msun) $\geq$ 13.5 bin, as there are no central galaxies with 9.7 $\leq$ $\log_{10}$(\Mstar/\Msun) $\leq$ 11.4 in the highest halo mass bin (see Fig. \ref{fig:mc200comp}). In \magc, we observe increasing suppression in SF with increasing \Mcrit in the outskirts for satellites, also similar to \eagle. At $R > 1.5$ \re, the profiles shift to lower ranges of \dssfr, with on average increasingly steeper negative gradients (-0.12 to -0.23 \slopeunits with increasing \Mcrit; see Table \ref{tab:ms_all_sims_slopes}) suggesting environmental processes at play in satellites. On the other hand, positive gradients of comparable slopes are captured within $R \leq 1.5$ \re across all \Mcrit bins, which may be partially explained by the comparable values of $\mathrm{f_{BH, THRESH}}$ (i.e., satellites with seeded BHs). 

The distinction between centrals and satellites is also clear for \tng. In comparison to the centrals, satellites show weak or negligible central suppression but a uniform decline from the centre to the outskirts, with profiles shifting to lower ranges of \dssfr with increasing \Mcrit. Strong central suppression comparable to \magc (1.39 vs. 1.04 \slopeunits) is captured in central galaxies residing in the intermediate \Mcrit bin. The values of $\mathrm{f_{BH, THRESH}}$, i.e. fraction of galaxies with BH masses above $M_{\rm BH} \geq$ 10\textsuperscript{8.2} \Msun, are significantly lower for satellites in comparison to centrals. This suggests that negligible SF suppression in the centre of \tng satellites is likely due to a significant portion of the population not being subject to AGN feedback. Similar to \magc, no profile is generated for centrals in the most massive halo bin due to low number statistics (only 3 galaxies). 

A clear trend with environment is only seen when the sample is constrained for both central/satellite status and \Mcrit, where these parameters together trace both internal and external processes. In our case, this trend was captured with central/satellite status and halo mass. Without splitting by halo mass, the median \dssfr trends between centrals and satellites are indistinguishable (i.e., black solid lines in each panel), which highlights the importance of the role of the environment in quenching galaxies, alongside internal processes such as AGN feedback. We also stress that these trends are discernible in this SF state due not only to population statistics, but also to the fact that galaxies on the SFMS have yet to be quenched on a global scale. We find that trends with environment are less conspicuous for galaxies below the SFMS for all simulations, as seen in the full set of \dssfr profiles in Appendix \ref{sec:app1}. This is likely due to the fact that after galaxies begin quenching on a global scale, both internal and external processes alike act throughout the entire galaxy disc, leading to similar radial trends regardless of central/satellite status and environment. 

\subsection{Comparison to MAGPI observations}
\label{subsec:magpi_comp}
We now explore how the simulations compare to the context of the MAGPI observations. Overall, the radial trends predicted by simulations are not in line with those of MAGPI observations, as shown in Fig. \ref{fig:allsimrprof}. Flat profiles are observed in star-forming galaxies above the SFMS for MAGPI, whereas negative gradients are observed for simulated galaxies in the same \dsfr bin. While MAGPI measures positive gradients for galaxies on and below the SFMS, suggestive of inside-out quenching, it is only for the lowest \dsfr bin (i.e., \dsfrq; red) where we see all three simulations agree with that picture. While \magc does exhibit a positive gradient within $R \sim 2$ \re in line with MAGPI for galaxies on and below the SFMS, the degree of suppression is stronger. 

While this is not the first time simulated galaxies on the SFMS have shown hints of outside-in quenching \citep[e.g.,][]{starkenburg2019}, it is still puzzling as to why this is the case. Under the assumption that environmental processes such as ram pressure stripping are indeed at play, outside-in quenching is an expected outcome as such stripping mechanisms are most effective on loosely bound gas in the outskirts \citep{gunngott1972}. In comparison to MAGPI, where the profiles are robustly measured out to $R \sim 3$ \re, we probe out to $R \sim$ 5 \re for the simulations. The use of different SFR indicators may also explain the discrepancy in part, where compared to instantaneous SFRs derived from simulations, \Halpha-based SFRs are not free from contamination due to non-SF diffuse ionised gas \citep[e.g.,][]{appleby2020}. Indeed, the MAGPI profile for galaxies on the SFMS is measured based on \Halpha. We also did note in \citetalias{mun2024} that the uptick in the outskirts at $R \gtrsim 2$ \re (see the \dsfrms panel in Fig. \ref{fig:allsimrprof}) is due to three highly star-forming (i.e., high \dsfr) galaxies with a significant number of \Halpha detections out to $R \sim$ 3 \re. Removing the three galaxies does result in a flatter trend albeit at the cost of dropping the last radial bin. However, given that the trend is still at odds with the negative gradients observed in simulations, this may indicate differences arising from feedback prescriptions and/or resolution. We also note the possibility that the negative gradients may be indicative of inside-out growth \citep[e.g.,][]{munoz-mateos07, vandokkum2010}. 

The profiles from simulations are on average normalised towards lower \dssfr for galaxies below the SFMS, relative to MAGPI observations, as shown in Fig. \ref{fig:allsimrprof}. The \dsfrgv (green) and \dsfrq (red) bins are where the contribution of D4000-based \sigsfr becomes dominant over \Halpha-based measurements in MAGPI galaxies. Due to limitations in the D4000-sSFR relation, there is a non-negligible fraction of non-detected spaxels in the central regions for both \dsfrgv and \dsfrq bins (see Appendix 3 and Fig. A5 in \citetalias{mun2024}). For the \dsfrgv bin, the distribution of non-detected spaxels is comparable to that of detected spaxels, such that the overall profile would be shifted to lower normalisations, in the event robust measurements replace upper limits. For the \dsfrq bin, there is a high fraction of non-detected spaxels at all radii but this fraction increases towards the centre, such that we expect the MAGPI profile to not only exhibit steeper gradients in \dssfr (hence stronger hints of inside-out quenching), but also be shifted towards lower normalisations. Simulations do not suffer from these detection limits, therefore differences in normalisations are expected. Taking this into account, there is better agreement between observations and simulations in the profiles in the lowest star-forming bin (i.e., \dsfrq), particularly with \magc. 

Above the SFMS, observational studies have found evidence of centrally enhanced SF at $z = 0$ which is attributed to starbursts triggered by mechanisms such as galaxy-galaxy interactions and mergers that preferentially funnel gas towards the centre \citep[e.g.,][]{ellison2018, ellison2020}. In contrast, at $z \sim 0.3$, MAGPI galaxies show on average a flat profile indicative of global enhancement, likely arising due to efficient gas mixing of a large gas reservoir. The simulations at $z \sim 0.3$ show centrally enhanced SF out to $\sim$3 \re. While this could be reflective of starbursts, it could also be a result of overcooling or differences in SF timescales.\footnote{All MAGPI galaxies in the \dsfrsb bin have \Halpha detections, where \Halpha traces timescales up to $\sim$10 Myr \citep{ke2012}. For the simulations, we are using instantaneous SFRs.} Overcooling refers to the gas that cools and collapses into stars too early, leading to galaxies with too concentrated SF and little angular momentum. The negative gradients measured for \eagle and \tng indicate centrally concentrated SF, in line with the expectations arising from the effects of overcooling. Differences in SF timescales also likely play a role, as positive gradients were previously observed for \tng galaxies at $z \sim 0$ when using time-averaged SFRs sensitive to 20 Myr timescales \citep{mcdonough2023}. 

Following the analysis done in Section \ref{subsec:rprof_sims}, we again break down the sample further to explore how some of the observed differences between MAGPI and the simulations may be ascribed to the different evolutionary processes impacting central and satellite galaxies. In each panel of Fig. \ref{fig:ms_all_sims_comp}, we overlay the median radial trends for centrals and satellites on the SFMS in MAGPI as black dotted lines. We also measure slopes for inner and outer regimes of each profile, again using $R = 1.5$ \re as the cutoff. The slopes and errors are presented in Table \ref{tab:ms_all_sims_slopes}. As mentioned in Section \ref{subsec:suppdata}, we are not able to measure the trends in different bins of \Mcrit for MAGPI in the same manner as done with the simulations due to the \Mcrit measurements being preliminary. 

As with the simulations, the median trends between the centrals and satellites are comparable for the MAGPI sample as well. The gradients measured in the inner regime (i.e., $R \leq 1.5$ \re) are similar, although we do measure notably steeper gradients beyond 1.5 \re for centrals. Given that positive gradients are measured in both the inner and outer regimes of both populations, this suggests that MAGPI centrals and satellites alike are subject to inside-out quenching. While MAGPI profiles are not robustly measured out to as far as $R > 4$ \re as with the simulations, there is nevertheless a clear discrepancy where all three simulations measure notable SF suppression in the outskirts. SF suppression in the centre is observed for \magc and \tng (positive gradients within $R \sim 1.5$ \re), with the strongest degree of suppression captured in centrals residing in higher mass halos. We discuss these implications further in Section \ref{subsec:disc_agn}. 

\begin{figure*}
    \centering
    \includegraphics[width=0.9\linewidth]{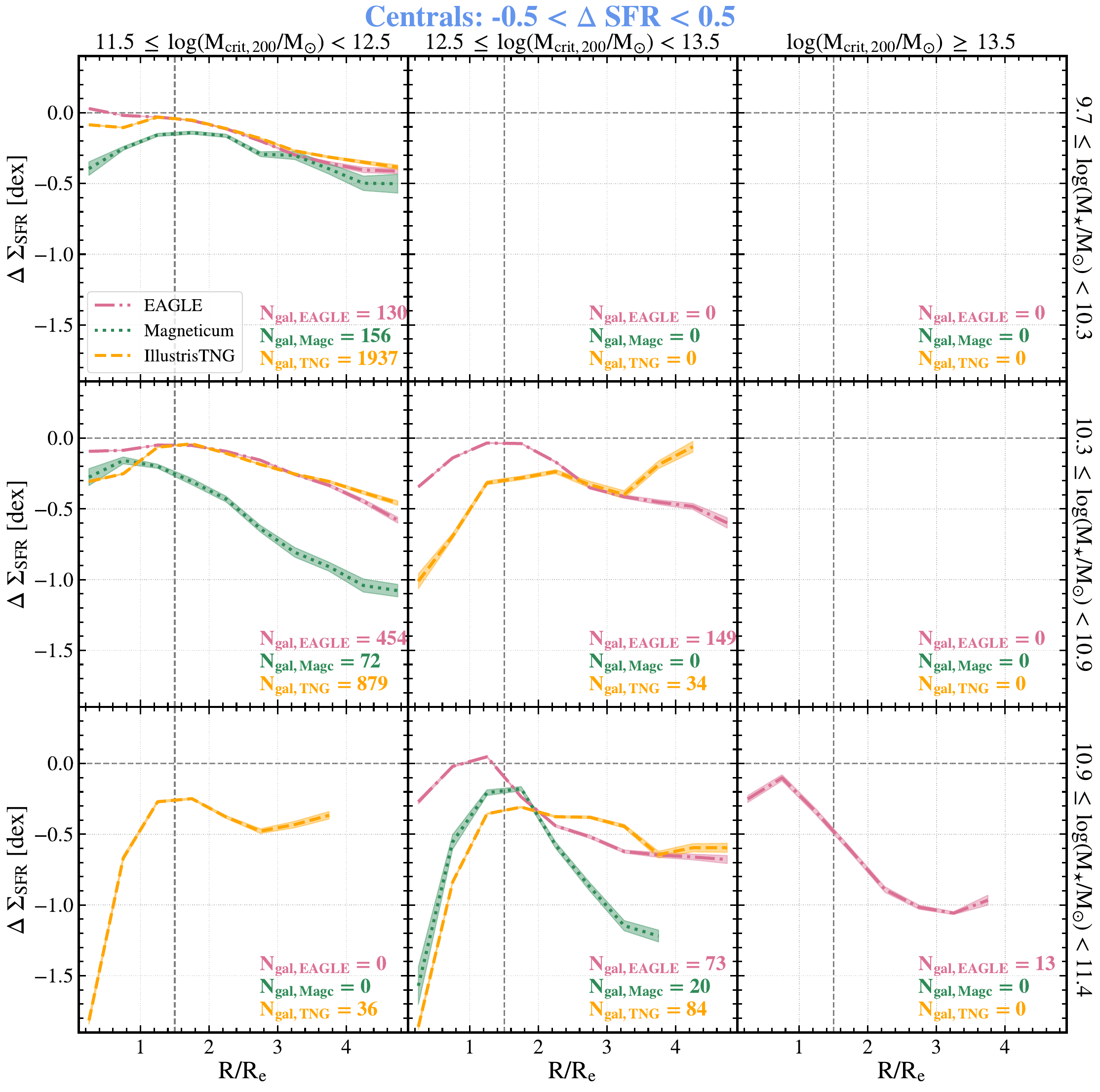}
    \caption{Median \dssfr profiles for central galaxies in the \dsfrms bin for \eagle (pink; dash-dotted), \magc (green; dotted), and \tng (orange; dashed) split across 6 different bins of \Mcrit and \Mstar. The same \Mcrit bins are used as done in Section \ref{sec:res} and Fig. \ref{fig:ms_all_sims_comp}. Each column indicates the \Mcrit bin and each row indicates the \Mstar bin. The total number of galaxies in each bin is denoted in the lower right corner of each panel, for each simulation in the same colours as the profiles. We observe a strong trend with \Mcrit and \Mstar across all three simulations. There is notable suppression in the centre ($R \leq 1.5$ \re) and the outskirts ($R > 1.5$ \re) with increasing \Mcrit, suggesting that centrals are subject to both internal and external processes.}
    \label{fig:ms_m200_mstar_cen}
\end{figure*}

\section{Discussion}
\label{sec:disc}
In the previous section, we found different behaviours of \dssfr profiles in the centres and outskirts of galaxies as a function of SF state, halo mass, and central/satellite classification. In this section, we discuss the role of AGN feedback and environment in shaping radial trends. By using mock observations in place of the raw simulation data, we remove a layer of discrepancy that inherently lies between observations and simulations. Most studies measuring radial trends have applied other means of matching to observations, such as emulating the sample selection method and measurement of radial profiles, and/or projecting galaxies in face-on orientations \citep{starkenburg2019, appleby2020, mcdonough2023}. We note that most of these studies measure sSFR profiles, which are slightly different from \dssfr, which is dependent on the locus of the resolved SFMS. Nevertheless, they are comparable given that both parameters are normalised by stellar mass. 

\subsection{The role of AGN feedback}
\label{subsec:disc_agn} 
We observe distinct radial profiles among \eagle, \magc, and \tng within the galaxy centre (i.e., $R \lesssim$ 1.5 \re), when we split the populations by centrals and satellites in different bins of \Mcrit (see Fig. \ref{fig:ms_all_sims_comp}). For example, the measured slopes within 1.5 \re differ by as much as 1.08 \slopeunits for centrals residing in halo masses of 12.5 $\leq$ $\log_{10}$(\Mcrit/\Msun) $<$ 13.5. We attribute these differences primarily to the implementation of AGN feedback, along with population differences that vary not only in stellar/BH masses, but also in the fraction of missing BHs. 

While all simulations roughly match the observed $z = 0$ stellar-to-BH mass relation, each simulation seeds BHs differently resulting in different BH mass distributions. For example, the BH seed mass in \tng is $\sim$10\textsuperscript{6} \Msun, whereas in \eagle it is $\sim$10\textsuperscript{5} \Msun. As a result, it is difficult to compare the role of BH using bins of BH mass in the same way as done for stellar mass and halo mass. 

As described in Section~\ref{sec:sims}, each simulation has its own unique prescription for AGN feedback, which determines its role in suppressing or even quenching star formation. As seen in the two far right panels of Fig. \ref{fig:allsimrprof}, the central uptick in \eagle galaxies below the SFMS is in agreement with previous \eagle-based results that suggest AGN feedback is insufficient in completely shutting down SF in the galaxy centre \citep{lagos2022}. In \eagle, the feedback is implemented as stochastic injection of thermal energy, which is distinct from the dual mode feedback recipes implemented in \magc and \tng. In Fig. \ref{fig:ms_all_sims_comp}, a stronger degree of central suppression is observed in \magc and \tng for centrals in the 12.5 $\leq \log_{10}$(\Mcrit/\Msun) $<$ 13.5 bin, where the measured gradients are steeper than \eagle by an average of 0.91 \slopeunits. The kinetic feedback mode in \tng has already been well-documented in the literature to be solely responsible for quenching massive galaxies inside-out \citep{weinberger2017, terrazas2020, nelson2021, hartley2023, kurinchi-vendhan2024}, especially in intermediate- to high-mass haloes ($\sim$10\textsuperscript{12} -- 10\textsuperscript{14} \Msun). Indeed, strong degrees of central suppression are observed for \tng galaxies below the SFMS in $\log_{10}$(\Mcrit/\Msun) $\geq$ 12.5 (see Fig. \ref{fig:gv_all_sims_comp} and \ref{fig:q_all_sims_comp}). It is also noteworthy that the kinetic mode replaced the bubble model for AGN in \textsc{Illustris} \citep{sijacki2007}, the latter which formerly failed to produce any central suppression in star-forming and green valley galaxies \citep{starkenburg2019}. 

\begin{figure*}
    \centering
    \includegraphics[width=0.9\linewidth]{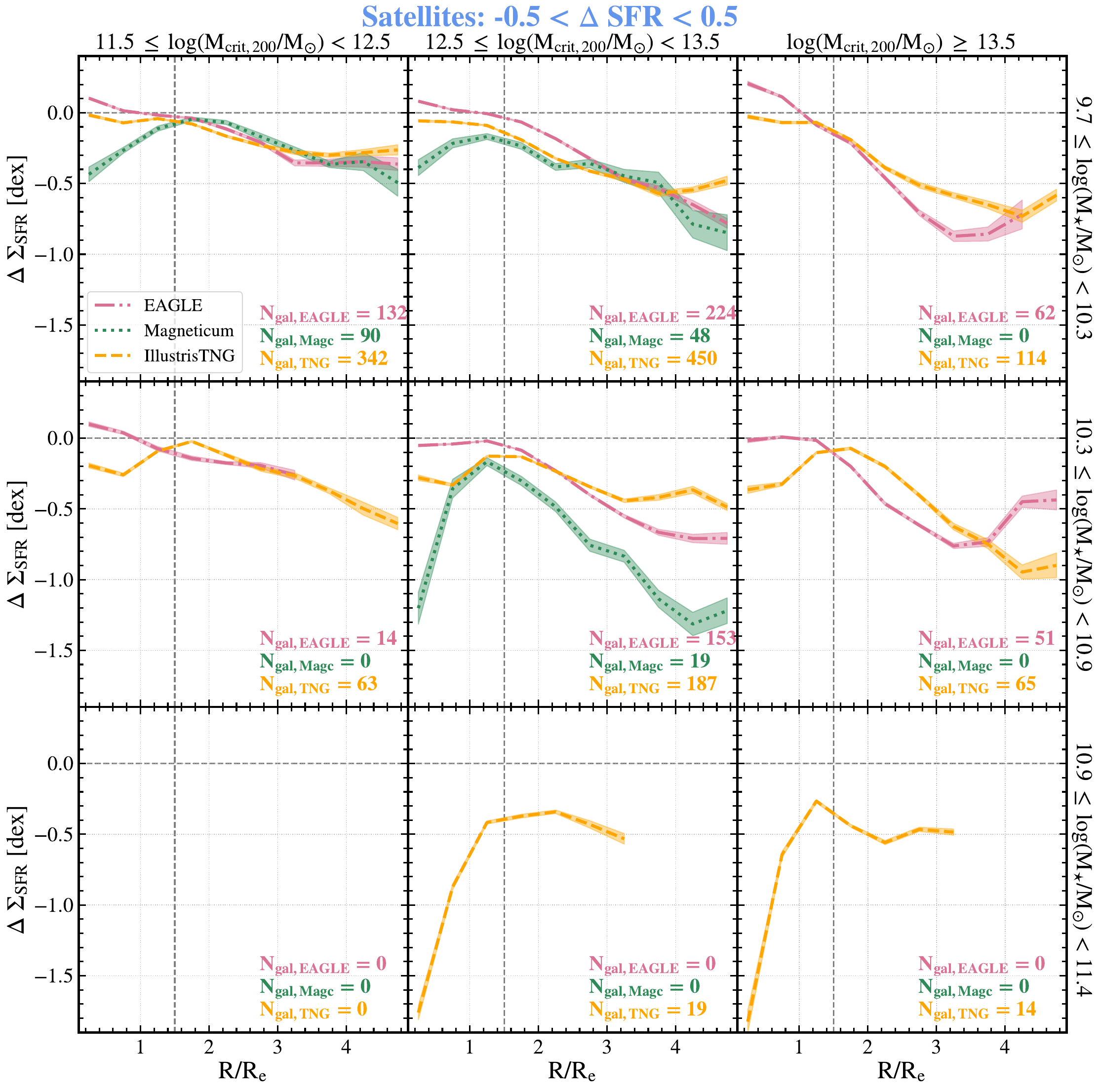}
    \caption{Analogous to Fig. \ref{fig:ms_m200_mstar_cen}, except for satellite galaxies in the \dsfrms bin. Satellites do not show as strong of a stellar mass dependence as centrals. High-mass satellites can experience central suppression likely due to AGN feedback, but due to low number statistics in the most massive (10.9 $\leq \log$[\Mstar/\Msun] $<$ 11.4) bin, we can only confirm a statistically significant trend with \tng. Overall, satellites exhibit negative gradients in the outskirts ($R > 1.5$ \re), indicative of predominantly environmental processes at play.}
    \label{fig:ms_m200_mstar_sat}
\end{figure*}

In \eagle, AGN feedback is known to be particularly effective for massive (\Mstar $>$ 3 $\times$ 10\textsuperscript{10} \Msun) central galaxies, where the interplay between the galaxy's hot halo gas and the BH plays a critical role in shutting down SF in the centre \citep{bower2017}. At lower masses, stellar feedback is particularly effective at preventing gas accretion into the BH. It is only at galaxy halo masses of 10\textsuperscript{11} -- 10\textsuperscript{12} \Msun where stellar feedback fails to prevent gas build up in the galaxy centre, leading to increasing BH accretion rates. This leads to hotter halos which then prevents cool gas inflow into the centre. For \tng, the kinetic mode feedback quenches galaxies with $M_{\rm BH}$ $\geq$ 10\textsuperscript{8.2} \Msun, which leads to a strong stellar mass dependence \citep{terrazas2020, piotrowska2022}. We can see this effect in the distribution of passive galaxies as a function of stellar mass in Fig. \ref{fig:sfmsdsfrsims}. For \tng, there is a notable over-density of passive galaxies at high stellar masses ($\gtrsim10^{10.5}$ \Msun) in contrast to a broader mass distribution seen in MAGPI and most other observational studies \citep[e.g.,][]{rp2015}. 

For EAGLE, there also seems to be a mass threshold at $\sim$10\textsuperscript{10} \Msun beyond which galaxies are preferentially quenched. However, this threshold is not built into the model; it arises from the interplay between the stellar and AGN feedback models. Conversely, the BH mass threshold in \tng exists by construction. Given these mass dependencies, the radial trends are likely to be affected by the fraction of galaxies with masses above these aforementioned thresholds, i.e., $\mathrm{f_{BH, THRESH}}$, in each simulation. As previously introduced in Section \ref{subsec:rprof_sims}, we find that $\mathrm{f_{BH, THRESH}}$ dramatically increases with \Mcrit for centrals in \eagle. For example, for galaxies on the SFMS, the fraction increases from $\sim$11 to $\sim$100 per cent in the highest \Mcrit bin. Despite this increase in the fraction of massive centrals, the degree of SF suppression within $R \sim 1.5$ \re is comparable across all \Mcrit bins, which is observed as well even when the profiles are split into different bins of stellar mass (see Fig. \ref{fig:ms_m200_mstar_cen}). 

We find a similar trend with \Mcrit for \tng as well, where the fraction of central galaxies with $M_{\rm BH} \geq$ 10\textsuperscript{8.2} \Msun increases from $\sim$4 to $\sim$90 per cent for galaxies on the SFMS. In contrast to \eagle, this translates into a strong trend with \Mcrit and \Mstar as seen in Fig. \ref{fig:ms_m200_mstar_cen} for \tng, where a stronger degree of SF suppression in $R \leq 1.5$ \re is captured with increasing \Mcrit and \Mstar. A similar increase is seen for galaxies in the \dsfrgv and \dsfrq bins, except with much higher $\mathrm{f_{BH, THRESH}}$ ($\sim$41 and $\sim$72 per cent, respectively) in the lowest \Mcrit bin, jumping to $\sim$100 per cent in the highest \Mcrit bin. The high fraction of massive central galaxies in the \dsfrq bin across all halo masses results in a similar degree of central suppression in SF, independent of halo mass (see Fig. \ref{fig:q_all_sims_comp}). For satellites, the fraction of massive galaxies does generally increase with \Mcrit in both \eagle and \tng, with the highest fractions measured for those in the \dsfrq bin. However, the fractions are much lower in comparison to those of the centrals, where there are less than $\sim$58 per cent of massive galaxies in any \Mcrit bin for both simulations. As such, satellites on the SFMS in both \eagle and \tng do not show as strong trends with \Mstar in fixed bins of \Mcrit (see Fig. \ref{fig:ms_m200_mstar_sat}), although massive satellites in \tng do show prominent dips in SF within $R \sim 1.5$ \re in all halo mass bins. 

We do not see a similar stellar/BH mass dependence of AGN feedback in \magc. This is likely due to the lack of a mass threshold at which the feedback mechanism is switched on. This is reflected in the passive population which evenly spans the full stellar mass range (Fig.~\ref{fig:sfmsdsfrsims}). However, there is a notable fraction of missing BHs in the satellite population. These are likely due to two possible scenarios, where galaxies may 1) lose their BHs due to the BHs not being pinned in the galaxy centre (as done in \eagle and \tng; see \citealt{crain2015} and \citealt{weinberger2017}), or 2) not have a BH as they did not have enough gas after surpassing the stellar mass threshold required for BH seeding \citep{hirschmann2014}. We check for all global SF states to find that there is an average of $\sim$21 per cent (i.e., $\mathrm{f_{BH, THRESH}} \sim$ 0.79) of satellites with missing BHs in each \Mcrit bin. The fraction of galaxies without BHs is independent of halo mass or star-forming state. These galaxies do not show a central suppression in their SF profiles since they cannot host AGN feedback. Given that this fraction does not constitute a majority in each subsample, notable central suppression is still observed for galaxies on and just below the SFMS (see Fig. \ref{fig:gv_all_sims_comp}), in contrast with \eagle and \tng. We also checked the fraction of central galaxies with missing BHs to find that it is either equivalent to zero or negligible ($\sim$4 per cent in each \Mcrit bin for galaxies on the SFMS). 

In the context of MAGPI observations, there is a clear mismatch with simulations in how well central suppression is reproduced in galaxies on (i.e., \dsfrms) and just below (i.e., \dsfrgv) the SFMS. Optical AGN spaxels were removed for MAGPI galaxies \textit{on} the SFMS to avoid contamination in \Halpha emission, whereas for galaxies \textit{below} the SFMS, D4000-based SFRs were measured for galaxies with \textit{only} composite/AGN spaxels and/or spaxels with weak emission lines. However, the fraction of galaxies with D4000-based SFRs measured for optical AGN spaxels is negligible ($\sim$1 per cent of the MAGPI sample), largely in part due to upper limits in the D4000-sSFR relation. Given difficulties in the consistent measurement of \Halpha-based SFRs for star-forming and AGN regions, we may potentially miss out on the effects of current AGN activity on SF for some galaxies. Nevertheless, central suppression is observed in galaxies on the SFMS, which may be due to either previous AGN activity that have long removed cold gas from the galaxy centre, misidentified AGN due to weak emission lines \citep[e.g.,][]{tozzi2023}, or missed AGN activity due to it peaking in other wavelength regimes \citep{padovani2017}. 

As discussed previously, the degree of central suppression in the simulations is directly linked to the potency of the AGN feedback prescription. The strongest SF suppression is captured for central galaxies in the densest environments as probed by halo mass, suggesting a potential link between environment and AGN feedback - which we discuss further in Section \ref{subsec:disc_env}. We are not able to reach the same conclusions for MAGPI, likely due to low number statistics compared to those of the simulations. 

A number of previous studies have also highlighted tensions between observations and simulations with regards to central suppression of SF. Locally, MaNGA results suggest green valley galaxies are subject to inside-out quenching \citep{belfiore2018}, just as we observe with MAGPI, yet \eagle, \textsc{Illustris}, and \textsc{SIMBA} simulations instead find strong suppression in the outskirts \citep{starkenburg2019, appleby2020}. When an additional mode of feedback is implemented in the \textsc{SIMBA} simulations, in the form of X-ray feedback, a central suppression is captured \citep{appleby2020}. However, even this decline in central SF is too steep compared to observed green valley galaxies in MaNGA, suggesting that AGN feedback may be more potent in \textsc{SIMBA} compared to other simulations \citep[e.g.,][]{wright2024}. 

An interesting exception to these findings is introduced by \citet{nelson2021}, where a remarkable agreement is captured in the radial trends between 3D-HST observations and TNG50 simulations at $z \sim 1$, the latter referring to the highest resolution run of \tng. Both samples measure inside-out quenching for galaxies on and below the SFMS, where central suppression is only captured in the TNG50 run with AGN kinetic feedback switched on, in line with previous \tng-based studies. \citet{nelson2021} partially attribute the success to building measurements of SFRs and stellar masses based on spectral energy distribution (SED) fitting with stellar population parameters derived via a Bayesian inference framework. The cell/particle data from TNG50 were also projected onto a grid to mimic the random orientation of observed galaxies, where they use the SFRs and stellar masses from the gas cells and star particles, respectively. This also suggests that differences in SFR indicators and/or timescales have a negligible impact on the radial trends, even for galaxies above the SFMS, where overcooling issues may be pronounced. The use of a higher resolution \tng run also likely minimizes such issues \citep{ss2016}. 

\subsection{The role of the environment}
\label{subsec:disc_env}
In our previous discussion of AGN feedback prescriptions, we found central suppression in SF to be more pronounced for central galaxies residing in higher-mass halos across all three simulations. Environmental effects are also observed in the galaxy outskirts, where increasing suppression of SF is captured with increasing halo mass. All of these observations suggest that the role of environment in quenching is multi-faceted. A study based on \tng galaxies at higher redshifts ($z \gtrsim 3$; \citealt{kurinchi-vendhan2024}) shows evidence for the interplay between environment and AGN feedback, where more massive halos provide a deeper gravitational potential well for the accelerated accretion of gas, fuelling BH growth and thus powering stronger AGN feedback. 

The role of environment in the outskirts of galaxies is most evident for satellites across all three simulations, where increasingly steeper negative gradients and offsets to lower \dssfr are measured with increasing halo mass, in comparison to the negligible changes observed in the centre. These trends are indicative of environmental processes that preferentially remove cold gas from the outskirts of galaxies, which include hydrodynamical and gravitational mechanisms. In \eagle, the most common mechanisms that deplete the cold gas content in satellites are ram pressure stripping due to the intragroup medium, tidal stripping due to the host halo, and interactions with other satellites \citep{marasco2016, bahe2017}. All of these mechanisms may easily disturb and/or strip the tenuous gas in the outskirts, quenching the galaxies outside-in. Findings based on \tng suggest that the dominant quenching mechanism for satellites depends on infall histories, where low-mass satellites may have quenched as early as 4 -- 10 Gyrs ago via pre-processing in earlier hosts \citep{donnari2021}. On the other hand, satellites undergoing strong degrees of ram pressure stripping (i.e., `jellyfish' galaxies) are typically late infallers ($<$2.5 -- 3 Gyrs with respect to $z = 0$; \citealt{yun2019}). Pre-processing is found to be dominant in low-mass satellites at $z = 0$, which may explain the weaker dependence (i.e., flatter gradients in $R > 1.5$ \re in comparison to \eagle and \magc) of the \dssfr trends on \Mcrit in \tng (see Fig. \ref{fig:ms_all_sims_comp}). Results from the larger \magc volumes of \textsc{box2} and \textsc{box2b} suggest that the orbital trajectories of galaxies in clusters determine their fates \citep{lotz2019}, where low-mass satellites travelling on radial orbits are more likely to experience ram pressure stripping. 

Most MAGPI galaxies used in this study reside in group environments. We thus do not expect environmental stripping mechanisms, such as ram pressure stripping and tidal interactions, to be as strong as in galaxy clusters \citep[e.g.,][]{marasco2016, donnari2021}. Furthermore, given that all MAGPI galaxies are observed within a projected radius of $\sim$270 kpc at $z \sim 0.3$ (typical group radius $\sim$1 -- 2 Mpc; \citealt{robotham2011}), most of them are likely located near the centre of the gravitational potential well of their host groups. Studies following orbital trajectories of infalling group/cluster galaxies in phase space suggest that once a galaxy has reached the core of a group/cluster, much of the star-forming gas has either been exhausted or stripped, such that the galaxy is globally quenched \citep[e.g.,][]{jaffe2015, rhee2017, yoon2017, barsanti2018}. As such, galaxies at the group/cluster centre are unlikely to show any signs of ongoing outside-in quenching. To corroborate these interpretations, we intend to explore the distribution of MAGPI and simulated galaxies in phase space in a future study. 

\section{Summary}
\label{sec:summ}
In this study, we explore what drives radial trends in star-formation profiles, at $z\sim0.3$, across different environments and star-forming states using data from the MAGPI survey and mock data cubes from three simulations -- \eagle, \magc, and \tng. Following \cite{mun2024}, we measure radial trends in \dssfr out to 5 \re using the global and resolved SFMS based on \sigsfr and \sigstar maps generated from mock observations for each simulation. A careful treatment of mock observations was made, including matching detection/resolution limits in SFR and \Mstar, matching parameter ranges of \Mstar and \Mcrit, and accounting for the average characteristics of MAGPI observations, including PSF and pixel scale. We focus on understanding the role of internal and external mechanisms in shaping radial trends of observed and simulated galaxies at $z \sim 0.3$. We summarize our key results below: 

\begin{itemize}
    \item The slope of the global SFMS at $z\sim0.3$ between MAGPI and all three simulations agree within 1 -- 2$\sigma$, whereas the slope of the resolved SFMS is shallower in the simulations than in the data. 
    \item When matched in stellar and halo mass ranges, the radial trends in simulations agree well with one another out to $\sim 4$ \re, with the exception of varying degrees of central suppression within $R \sim 1.5$ \re for galaxies on and below the SFMS, due to different AGN feedback prescriptions. 
    \item The role of environment is explored using central/satellite classification and halo mass. Different behaviour between centrals and satellites is only seen when also split into different \Mcrit bins. This points to the subtle role of environment and highlights the need for large samples such that important trends are not missed or averaged out when exploring radial profiles.
    \item In general, central galaxies experience an increasing degree of central ($< 1.5$ \re) suppression in SF with increasing halo and stellar mass, due to the onset of AGN feedback. Both central and satellites are susceptible to environmental mechanisms acting on the outskirts ($>1.5$ \re), which include ram pressure stripping and tidal interactions with other galaxies and/or the group/cluster environment. High-mass satellites can also experience prominent suppression in the centre, although this is likely simulation dependent. The halo mass dependence of suppression at the outskirts is most prominent at large radii ($\sim3$ \re), a difference that is difficult to observe with fixed integral field unit (IFU) sizes.
    \item When comparing the simulations to the MAGPI observations at $z\sim0.3$, we only observe a strong qualitative agreement in radial trends for galaxies far below the SFMS, where all four samples show signs of inside-out quenching. These results suggest that the prescriptions behind the quenching mechanisms, such as AGN feedback, may be insufficient in accurately reproducing the degree of suppression in quenching galaxies with residual SF. Some of the discrepancies with MAGPI observations may arise not only due to differences in number statistics, but also the environments probed. As most MAGPI galaxies are found near the centre of group environments in projection, galaxies below the SFMS are unlikely to show signs of outside-in quenching. In a future study, we plan to study the distribution of MAGPI and simulated galaxies in phase space. 
\end{itemize}

While reproducing global galaxy trends has long been a target of large hydrodynamical simulations, spatially resolved properties, like radial profiles, provide an additional test of the underlying models. The subtle differences shown in this work as a function of radius suggest that the different pathways to reproducing the local diversity of galaxy properties \citep{foster2021} may be a direct result of AGN feedback implementation.

The findings from \cite{mun2024} and this work emphasize the importance of radial profiles that extend well beyond 1 \re. This is difficult for most studies as it requires large number statistics, deep observations, along with IFUs with FOVs large enough to probe the full extent of the large scale environment. Future instrumentation concepts such as the Wide Field Spectroscopic Telescope (WST) would be well suited to extend this work. 

\section*{Acknowledgements}
We wish to thank the ESO staff, and in particular the staff at Paranal Observatory, for carrying out the MAGPI observations. MAGPI targets were selected from GAMA. GAMA is a joint European-Australasian project based around a spectroscopic campaign using the Anglo-Australian Telescope. GAMA was funded by the STFC (UK), the ARC (Australia), the AAO, and the participating institutions. GAMA photometry is based on observations made with ESO Telescopes at the La Silla Paranal Observatory under programme ID 179.A-2004, ID 177.A-3016. The MAGPI team acknowledge support by the Australian Research Council Centre of Excellence for All Sky Astrophysics in 3 Dimensions (ASTRO 3D), through project number CE170100013. We acknowledge the Virgo Consortium for making their simulation data available. The EAGLE simulations were performed using the DiRAC-2 facility at Durham, managed by the ICC, and the PRACE facility, Curie, based in France at TGCC, CEA, Bruy\'{e}res-le-Ch\^{a}tel. We thank the \tng team for making their simulation data available. KH acknowledges funding from the Australian Research Council (ARC) Discovery Project DP210101945. CL, JTM and CF are the recipients of ARC Discovery Project DP210101945. CF is the recipient of an Australian Research Council Future Fellowship (project number FT210100168) funded by the Australian Government. LMV acknowledges support by the German Academic Scholarship Foundation (Studienstiftung des deutschen Volkes) and the Marianne-Plehn-Program of the Elite Network of Bavaria. MB acknowledges the support of McMaster University through the William and Caroline Herschel Fellowship. TG acknowledges support from ARC Discovery Project DP210101945. YM is supported by an Australian Government Research Training Program (RTP) Scholarship. SMS acknowledges funding from the Australian Research Council (DE220100003). 

%%%%%%%%%%%%%%%%%%%%%%%%%%%%%%%%%%%%%%%%%%%%%%%%%%
\section*{Data Availability}
All MUSE data present in this work are publicly available on the \href{http://archive.eso.org/cms.html}{ESO Science Archive Facility}. Data products such as fully reduced datacubes, emission line fits, and mock data cubes from simulations will be made available as part of a MAGPI team data release (Mendel et al. in preparation; Battisti et al. in preparation; Harborne et al. in preparation). 

The \eagle simulations are publicly available; see \citet{mcalpine2016, eagle2017} for how to access \eagle data. \tng data is publicly available from \url{https://www.tng-project.org} \citep{nelson2019}. Access to \magc data can be given upon request (\url{http://www.magneticum.org/faq.html#DATA}).


\bibliographystyle{mnras}
\bibliography{biblio} 

%%%%%%%%%%%%%%%%%%%%%%%%%%%%%%%%%%%%%%%%%%%%%%%%%%

%%%%%%%%%%%%%%%%% APPENDICES %%%%%%%%%%%%%%%%%%%%%

\appendix

\section{Central \& satellite \texorpdfstring{$\boldsymbol{\mathrm{\Delta \Sigma_{SFR}}}$}{dss} profiles}
\label{sec:app1}

\begin{figure*}
    \centering
    \includegraphics[width=0.85\linewidth]{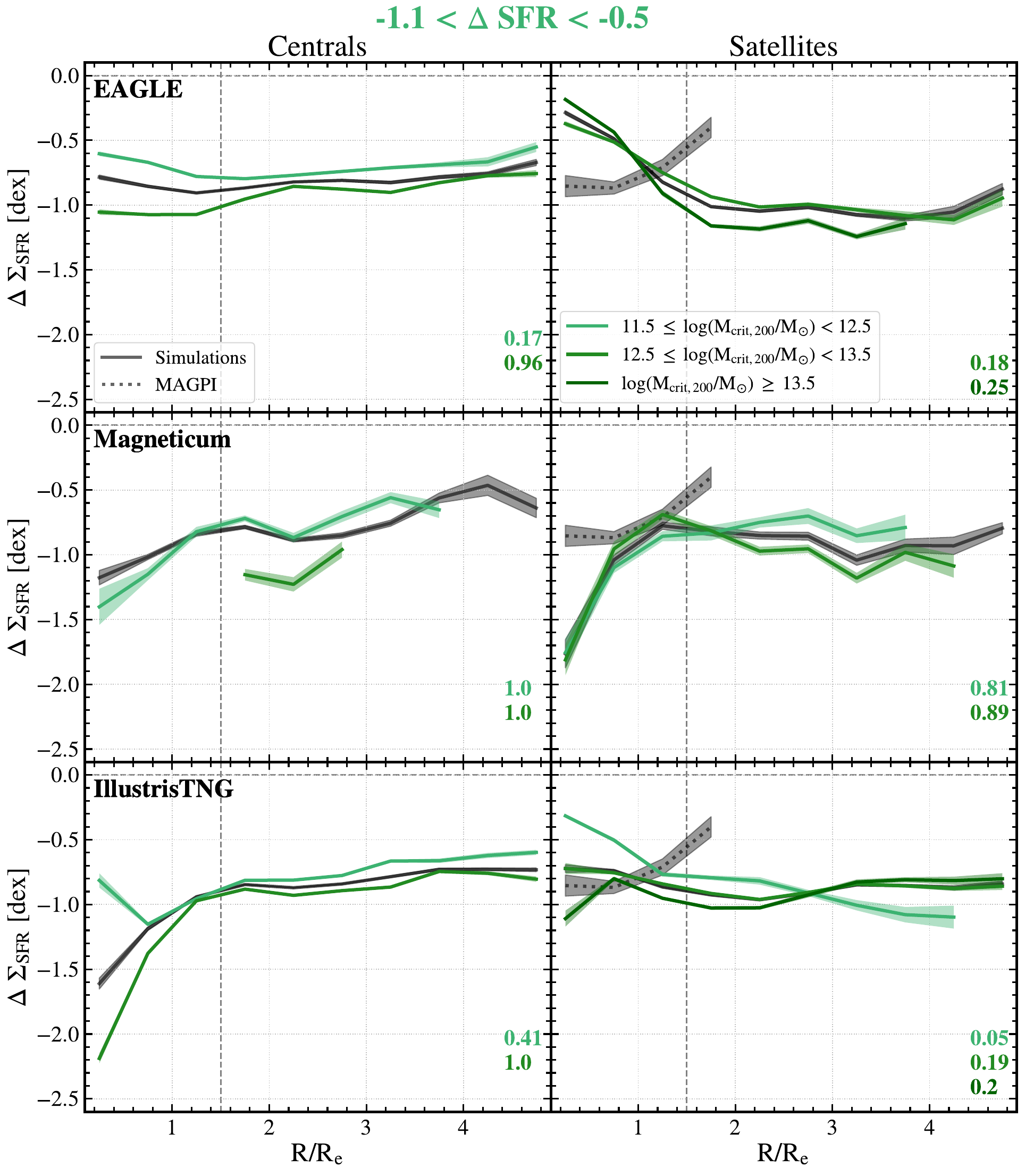}
    \caption{\dssfr profiles of for galaxies just below the SFMS (i.e., \dsfrgv) shown separately for centrals (left column) and satellites (right column), for MAGPI and all three simulations. This figure is analogous to Fig. \ref{fig:ms_all_sims_comp}.}
    \label{fig:gv_all_sims_comp}
\end{figure*}

\begin{figure*}
    \centering
    \includegraphics[width=0.85\linewidth]{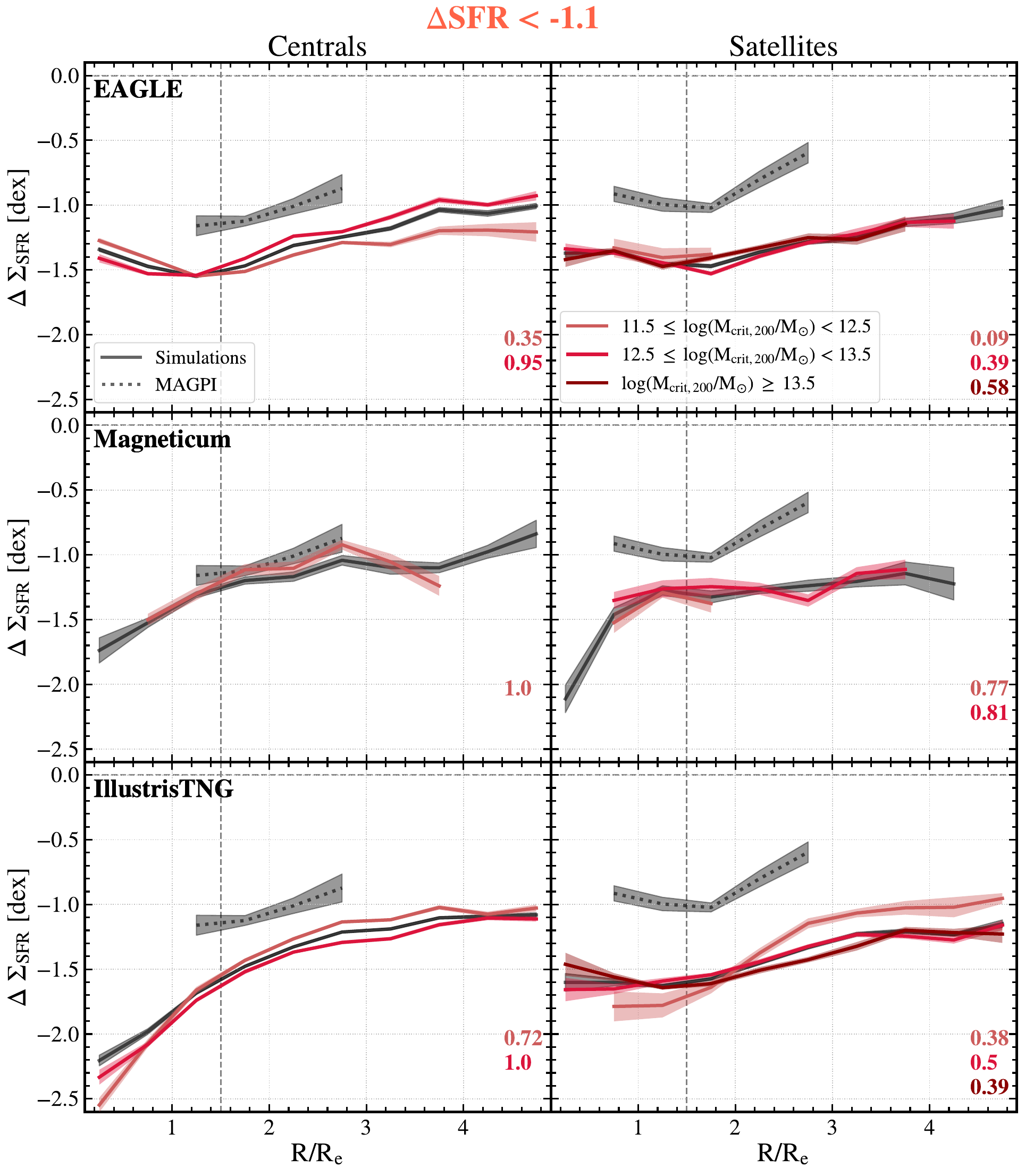}
    \caption{Analogous to Fig. \ref{fig:ms_all_sims_comp} and \ref{fig:gv_all_sims_comp}, showing \dssfr profiles for galaxies below the SFMS (i.e., \dsfrq).}
    \label{fig:q_all_sims_comp}
\end{figure*}

Following Section \ref{sec:res}, we show the full set of \dssfr profiles for centrals and satellites from each simulation split into different \Mcrit bins, as a function of global SF states. Profiles for the \dsfrsb bin are not shown due to low number statistics in both centrals and satellites, making fair comparisons between the two across different bins of \Mcrit infeasible. In the two \dsfr bins shown here, we find a weaker environmental dependence of \dssfr trends as galaxies begin quenching on global scales. For the \dsfrgv (Fig. \ref{fig:gv_all_sims_comp}) bin, the profiles between centrals and satellites are still distinguishable, with centrals on average showing stronger suppression in the centre with increasing halo mass. However, for galaxies in the \dsfrq (Fig. \ref{fig:q_all_sims_comp}) bin, aside from \tng, the profiles overlap with one another across all bins of \Mcrit, with comparable radial profile shapes for centrals and satellites. 

\section{\texorpdfstring{$\boldsymbol{\mathrm{\Delta \Sigma_{SFR}}}$}{dss} profiles with percentiles as errors}
\label{sec:app3}

Following the discussion in Section \ref{subsec:rprof}, we show identical \dssfr profiles as those in Fig. \ref{fig:allsimrprof}, except with the 25th and 75th percentiles shown in place of the bootstrap errors on the median. We observe that there is a large spread in the individual trends of galaxies in each \dsfr bin, where there is even significant overlap in the range of \dssfr values across multiple \dsfr bins. 

\begin{figure*}
    \centering
    \includegraphics[width=0.9\linewidth]{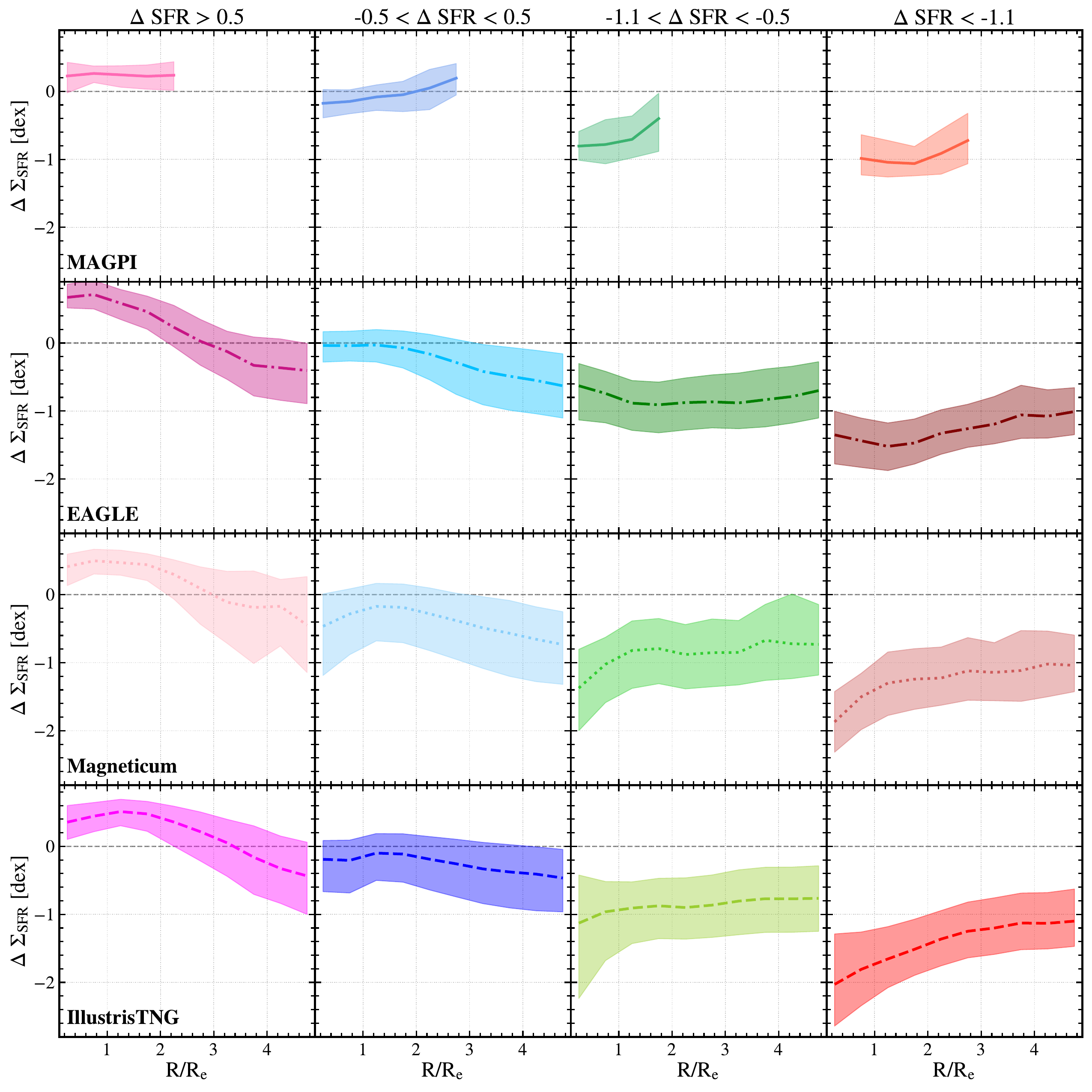}
    \caption{Top to bottom: Median \dssfr profiles for MAGPI, \eagle, \magc, and \tng for each global SF state (in order of decreasing \dsfr, from left to right). These are identical to those in Fig. \ref{fig:allsimrprof}, except with the 25th and 75th percentiles shown instead of bootstrap errors, to show the spread of the data embedded in the profiles.}
    \label{fig:rprof_percent}
\end{figure*}

%%%%%%%%%%%%%%%%%%%%%%%%%%%%%%%%%%%%%%%%%%%%%%%%%%

% Don't change these lines
\bsp	% typesetting comment
\label{lastpage}
\end{document}